\documentclass[10pt,twocolumn,showpacs,preprintnumbers,amsmath,amssymb,aps,prl,longbibliography,superscriptaddress,nofootinbib]{revtex4-2}
\usepackage{array,braket,mathtools,siunitx}

\usepackage[unicode=true,
bookmarks=true,bookmarksnumbered=true,bookmarksopen=false,
breaklinks=true,pdfborder={0 0 0},backref=false,colorlinks=true]{hyperref}
\hypersetup{citecolor={blue},urlcolor={magenta}}

\usepackage{cleveref}
\crefname{appendix}{App.}{Apps.}
\crefname{equation}{Eq.}{Eqs.}
\crefname{figure}{Fig.}{Figs.}
\crefname{table}{Tab.}{Tabs.}
\crefname{section}{Sec.}{Secs.}

\begin{document}
\title{Inhomogeneous Quantum Quenches of Conformal Field Theory with Boundaries
}

\author{Xinyu Liu}
\affiliation{Department of Physics, Princeton University, Princeton, New Jersey 08544, USA}
\affiliation{Department of Physics, California Institute of Technology,
Pasadena, California 91125, USA}
\author{Alexander McDonald}
\affiliation{Department of Physics, Princeton University, Princeton, New Jersey 08544, USA}
\author{Tokiro Numasawa}
\affiliation{Institute for Solid State Physics, University of Tokyo, Kashiwa 277-8581, Japan}
\author{Biao Lian}
\affiliation{Department of Physics, Princeton University, Princeton, New Jersey 08544, USA}
\author{Shinsei Ryu}
\affiliation{Department of Physics, Princeton University, Princeton, New Jersey 08544, USA}
\date{\today}

\begin{abstract}
We develop a method to calculate generic time-dependent correlation functions for inhomogeneous quantum quenches in (1+1)-dimensional conformal field theory (CFT) induced by sudden Hamiltonian deformations that modulate the energy density inhomogeneously. Our work particularly focuses on the effects of spatial boundaries, which have remained unresolved by previous analytical methods. For generic post-quench Hamiltonian, we develop a generic method to calculate the correlations by mirroring the system, which otherwise are Euclidean path integrals in complicated spacetime geometries difficult to calculate. 
On the other hand, for a special class of inhomogeneous post-quench Hamiltonians, including the M\"obius and sine-square-deformation Hamiltonians, 
we show that the quantum quenches exhibit simple boundary effects calculable from Euclidean path integrals in a straightforward strip spacetime geometry. 
Applying our method to the time evolution of entanglement entropy, we find that for generic cases, the entanglement entropy shows discontinuities (shockwave fronts) propagating from the boundaries.
In contrast, such discontinuities are absent in cases with simple boundary effects. We verify that our generic CFT formula 
matches well with numerical calculations from free fermion tight-binding models for various quench scenarios.
\end{abstract}

\maketitle

Recent developments in quantum devices and simulators enable us to explore 
far-from-equilibrium quantum dynamics of many-body systems, which is of both fundamental and practical importance. A paradigm is quantum quenches which have been studied extensively both
theoretically and experimentally \cite{Calabrese_2006, Calabrese_2007, Calabrese_2007b, Calabrese_2016,  Kaufman_2016,  Wen_2018,  goto2021nonequilibrating,  2005JSMTE..04..010C,  2019Sci...364..256L, 2019Sci...364..260B,  PhysRevLett.109.020504,  Alcaraz_2011,  Islam_2015,  Nozaki_2014,  Sotiriadis_2008,  Horvath_2022,Capizzi_2023,doi:10.1073/pnas.1703516114} 
where a system is initially prepared in a stationary state of some Hamiltonian
and then time-evolved by another Hamiltonian.
The post-quench evolution of entanglement entropy and other quantities can reflect intrinsic dynamical properties of many-body systems such as ergodicity/non-ergodicity. 
Particularly, quantum quenches with inhomogeneity from disorder 
or intentional modulation yield
intriguing dynamics.
For instance, 
quantum quench by a (1+1)d inhomogenous Hamiltonian 
with couplings with square-root dependence on    
site indices,  
dubbed square-root deformation (SRD) in
\cite{Wen_2016}, 
allows perfect distant quantum communication \cite{perfect2004};
For (1+1)d quantum many-body systems at 
conformal criticality, 
quantum quench by the so-called 
M\"obius and sine-square deformation (SSD) 
can lead to heating/non-heating dynamics and create a black-hole like excitation 
\cite{shinsei, goto2023scrambling}.

Although important, analytical solutions to quantum quench problems in interacting many-body systems are still rare. Nonetheless, 
previous studies have demonstrated that a wide class of inhomogeneous quantum quench and Floquet problems in $(1+1)$d conformal field theory (CFT) are amendable to exact solutions -- see, for example,
\cite{2018JSP...172..353G,
PhysRevLett.122.020201,
Wen_2018,
wen2018floquet,
Fan_2020,
Wen_2021,
Fan_2021,
Wen_2022,
Han_2020,
Caputa_2014,
Lapierre_2020a,Lapierre_2020b,
Moosavi_2021, Lapierre_2021, 
goto2023scrambling,
2022arXiv221204201D,
2022JHEP...08..221D}. 
In particular, for finite systems with periodic boundary conditions (PBC) \cite{goto2023scrambling, Moosavi_2021, Lapierre_2021, Fan_2020}, 
analytical formulas for generic smooth inhomogeneous quenches are derived, which reveal rich physics of inhomogeneous CFTs. 
On the other hand, for finite systems with 
spatial boundaries,
analytical solutions are only found for
specific types of quenches 
such as M\"obius and SSD quenches \cite{Lapierre_2021, Fan_2020}, while the generic inhomogeneous quench problems suffer from boundary effects and have not been solved.

In this letter, we study 
the time-dependent correlation 
functions and 
entanglement entropy of 
a wide class of $(1+1)$d inhomogeneous CFT quenches with 
boundaries.
As we will show,
these processes can be 
represented as 
Euclidean path integrals in complicated spacetime geometries.
While the path integrals are difficult to perform in general,  
we develop a method circumventing this difficulty and derive 
an exact formula \cref{Ox-generic} characterizing the \emph{generic boundary effect}. 
We also show that a special class of 
quench problems
with boundaries, 
including M\"obius and SSD quenches studied previously \cite{Lapierre_2021, Fan_2020}, reduce to Euclidean path integrals in a simple strip geometry, and the generic \cref{Ox-generic} reduces to \emph{simple boundary effect} in \cref{Ox-simple}. 
For the case of generic boundary effect, physical observables such as entanglement entropy typically exhibit a shock-wave behavior propagating from the boundaries, which is absent for the case of simple boundary effect.
We verify that our generic formula matches well with free fermion tight-binding calculations for various quench problems in \cref{tbl:fx}. 

\indent\emph{Setup}. 
We consider a (1+1)d CFT with central charge $c$. 
The class of conformal quench problems of interest is defined on a finite spatial interval $[0, L]$, bounded by spatial boundaries. These boundaries completely reflect incoming energy flux by gluing the left- and right-moving components of the CFT energy-momentum tensor, thereby preventing energy flux through the boundaries
\cite{cardy2008boundaryconformalfieldtheory}.
This does not completely specify the boundary condition, since there is commonly more than one conformally invariant boundary condition in CFT. 
However, our method presented below does not depend on the specific type of conformally invariant boundary condition.
We also note that
the case of (semi)infinite interval can be derived by taking the limit $L\to \infty$. Initially, 
we assume the system is in the (unique) ground state $|\psi_0\rangle$ of the uniform CFT Hamiltonian 
\begin{equation}\label{eq-H0}
H_0=\int_0^L h(x)dx \ ,
\end{equation}
where $h(x)$ is the Hamiltonian density operator. 
Starting from time $t=0$, the Hamiltonian is suddenly changed to 
\begin{equation}
\label{def}
H=\int_0^Lf(x)h(x)dx\ , 
\end{equation}
where $f(x)$ is an arbitrary smooth real non-negative function for $x\in[0,L]$. For lattice many-body systems, \cref{eq-H0,def} correspond to discrete lattice Hamiltonians of the form $H_0=\sum_x h_x$ and $H=\sum_x f(x)h_x$.  
The deformed Hamiltonians of this type 
have been studied 
for various choices of $f(x)$ 
\cite{
Vitagliano_2010,
Ram_rez_2014,
Ram_rez_2015,
Rodr_guez_Laguna_2017, 
2009PThPh.122..953G,
2011PhRvA..83e2118G,
PhysRevB.83.060414,
2012JPhA...45k5003K,
2015JPhA...48E5402I,
2016IJMPA..3150170I,
2016arXiv160309543O,
2018PTEP.2018f1B01T,
Cardy_2016,
Dubail_2017,
Wen_2016, 
MacCormack_2019}. 
Our goal is to calculate time-dependent correlation functions 
  after the quench.
  Our approach below in principle allows us to calculate
  arbitrary $k$-point correlation functions. 
  Explicitly, we demonstrate our method by studying the 
  entanglement entropy that can be expressed with one-point correlation function.

Let us start by observing that, since the Hamiltonian generates time translation, the quench
from $H_0$ to $H$ can be viewed as a sudden change of the metric from $ds_0^2$ to $ds^2$ at time $t=0$:
\begin{equation}\label{eq:metric}
ds_0^2=-d\tilde{t}^2+d\tilde{x}^2 \quad  \rightarrow \quad ds^2=-f(x)^2dt^2+dx^2\ .
\end{equation}
Here we use $(\tilde{x},\tilde{t})$ and $(x,t)$ to denote the spacetime coordinates before and after time $t=0$, respectively, with $\tilde{x}=x$ at $t=0$. This amounts to a change in the speed of light from $1$ to $f(x)$ at $t=0$. 

As the first step, 
we redefine the post-quench spatial coordinate $x$ into a coordinate $y$:
\begin{equation}\label{y(x)}
\frac{d y}{d x}=\frac{1}{f(x)}, \quad y=\int^x \frac{d x^{\prime}}{f\left(x^{\prime}\right)}=g(x)\ ,
\end{equation}
in which the post-quench speed of light is normalized back to $1$. Particularly, the spatial boundary points are mapped to $y_L= g(0)$ and $y_R=g(L)$.

To extend the coordinate $y$ into pre-quench times, we switch to Euclidean time $\tilde{t}=-i\tilde{\tau}$ and $t=-i\tau$, and assume $f(x)$ and thus $g(x)$ can be analytically continued into the complex plane (which would be true if $f(x)$ is real analytic). For the pre-quench spacetime, we perform a conformal transformation from complex coordinate $z=\tilde{x}+i\tilde{\tau}$ to $w=y+i\tau=g(z)$ (thus $dw/dz=f(z)^{-1}$). The metric $ds_0^2$ ($ds^2$) before (after) quench in \cref{eq:metric} can then be rewritten in coordinates $(y,-i\tau)$ as
\begin{equation}\label{eq:metric-cft}
ds_0^2=|f(z)|^2\left(d\tau^2+dy^2\right)\ \rightarrow\  ds^2=f(x)^2\left(d\tau^2+dy^2\right)
\end{equation}
Thus, the coordinate $w=y+i\tau$ glues together the metrics $ds_{0}^{2}$ and $ds^{2}$ at $t=0$ without changing the speed of light. 

\begin{figure}[tbp]
\begin{center}
\includegraphics[width=3.3in]{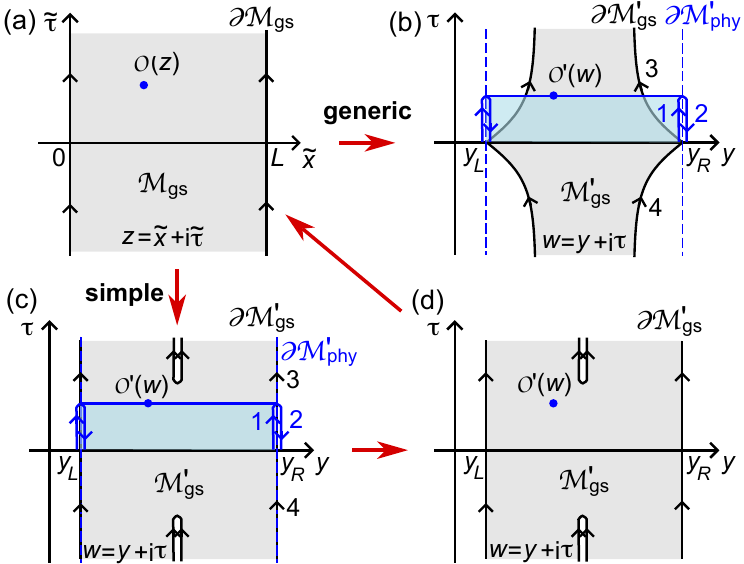}
\end{center}
\caption{(a) The initial state path integral manifold $\mathcal{M}_\text{gs}$ in $z$ coordinate. (b) Path integral representation ($1\rightarrow 2\rightarrow 3\rightarrow 4$). 
(c) Panel (b) for the M\"obius quench, which can be reduced to (d). 
}
\label{fig1}
\end{figure}

We denote operators in pre-(post-)quench coordinate $z$ ($w$) as unprimed (primed). We are interested in $k$-point correlation functions of nonchiral local operators $\mathcal{O}_j'$ at post-quench coordinates $w_j=y_j+i\tau$ at equal time $\tau$:
\begin{equation}\label{O(w)}
\begin{split}
&  \langle\prod_{j=1}^k\mathcal{O}_j^{\prime}(w_j)\rangle =\mathrm{tr}\Big[|\psi_{0}\rangle\langle\psi_{0}|
  e^{H\tau}\prod_{j=1}^k\mathcal{O}_j^{\prime}(y_j)
  e^{-H\tau}\Big] .
\end{split}
\end{equation}
This can be expressed as a glued path integral as follows. First, the ket (bra) ground state $|\psi_0\rangle$ ($\langle \psi_0|$) of $H_0$ can be obtained by an Euclidean-time path integral in the half-infinite strip $\mathcal{M}_{\text{gs}}^<$ ($\mathcal{M}_{\text{gs}}^>$) in the pre-quench coordinate $z=\tilde{x}+i\tilde{\tau}$, defined as the $\widetilde{\tau}<0$ ($\widetilde{\tau}>0$) region of strip $\mathcal{M}_{\text{gs}}$ (manifold $\text{Re}(z)\in[0,L]$) in \cref{fig1}(a). When mapped into the post-quench coordinate $w$, $|\psi_0\rangle$ ($\langle \psi_0|$) becomes a path integral over $\mathcal{M}_{\text{gs}}^{\prime <}$ ($\mathcal{M}_{\text{gs}}^{\prime >}$), which denotes the $\tau<0$ ($\tau>0$) region of manifold $\mathcal{M}_{\text{gs}}^{\prime}$ (grey region in \cref{fig1}(b)) that is the image of strip $\mathcal{M}_{\text{gs}}$ from $z$ to $w$. Second, the evolution operator $e^{-H\tau}$ can be expressed as a path integral in coordinate $w$ in the blue rectangle ranging from time $0$ to $\tau$ in \cref{fig1}(b). In sum, we can rewrite \cref{O(w)} as a path integral in a manifold given by $\mathcal{M}_\text{gs}'$ glued with two rectangles in the order $1\rightarrow2\rightarrow3\rightarrow4$ as shown in \cref{fig1}(b), with the operators $\mathcal{O}_j'$ inserted between rectangles $1$ and $2$, which has a complicated spacetime geometry (see SM \cite{suppl} Sec.\ I for details).

\emph{Simple boundary effect.} In the $w$ coordinate, the post-quench spacetime boundary are the $\tau$-independent lines at $y_L= g(0)$ and $y_R=g(L)$ (dashed lines in \cref{fig1}(b)), which we denote as $\partial {\cal M}'_{{\rm phy}}$. There exists a simple class of functions $f(x)$, for which the two boundaries $\partial\mathcal{M}_{\text{phy}}^{\prime}$ and $\partial\mathcal{M}_{\text{gs}}^{\prime}$ (boundary of $\mathcal{M}_{\text{gs}}^{\prime}$) exactly match within a finite Euclidean time interval $(-\tau_0,\tau_0)$ with some $\tau_0>0$. This includes the case $y_L=-\infty$ ($y_R=\infty$), where $\partial\mathcal{M}_{\text{phy}}^{\prime}$ and $\partial\mathcal{M}_{\text{gs}}^{\prime}$ around $y_L$ ($y_R$) are infinitely away and effectively match. An example is the previously studied M\"obius quench
(\cref{tbl:fx}) shown in \cref{fig1}(c). In this case, for $\tau\in(-\tau_0,\tau_0)$, the path integral in rectangle $2$ cancels with a part of $3$, thus the full path integral reduces to a path integral simply in the manifold $\mathcal{M}_\text{gs}'$ in \cref{fig1}(d). The conformal symmetry then relates the $k$-point function \cref{O(w)} in $w$ coordinates in $\mathcal{M}_\text{gs}'$ with the $k$-point function in $z$ coordinates in a strip $\mathcal{M}_\text{gs}$ in \cref{fig1}(a) as $\langle\prod_{j=1}^k\mathcal{O}_j^{\prime}(w_j)\rangle_{\mathcal{M}_\text{gs}'}=\prod_{j=1}^{k}\left|\frac{dw_j}{dz_j}\right|^{-\Delta_j} \langle\prod_{j=1}^k\mathcal{O}_j(z_j)\rangle_{\mathcal{M}_\text{gs}}$, where $z_j=g^{-1}(w_j)$, and $\Delta_j$ is the scaling dimension of $\mathcal{O}_j^{\prime}$ which maps to $\mathcal{O}_j$ in $z$ coordinate. In particular, this gives one-point function of an operator $\mathcal{O}'$ with scaling dimension $\Delta$ at position $x=g^{-1}(y)\in[0,L]$ and post-quench real time $t\ge 0$ as (SM \cite{suppl} Sec.\ II)
\begin{equation}\label{Ox-simple}
\langle\mathcal{O}^{\prime}(y,t)\rangle=\left[ \left|\frac{\partial x_+}{\partial x}\frac{\partial x_-}{\partial x}\right|^{-\frac{1}{2}}\frac{2 L}{\pi\epsilon} \sin \frac{\pi \left(x_++x_-\right)}{2L}\right]^{-\Delta},
\end{equation}
where $x_\pm = g^{-1}(y_\pm)= g^{-1}(g(x) \mp t)$ (\emph{not} restricted in $[0,L]$) are the light cone coordinates, which are the initial positions of a right-/left-moving quasiparticle reaching position $x$ at time $t$. As shown in SM \cite{suppl} Sec.\ IV, $g^{-1}(y)$ (thus $x_\pm$) in this case has analytical continuation on the entire real axis $y\in\mathbb{R}$, thus \cref{Ox-simple} holds for all $t\ge0$. 

We say the above class of 
quenches
have \emph{simple boundary effect}, which is easy to calculate due to the simple geometry of strip $\mathcal{M}_\text{gs}$. Particularly, if $\partial\mathcal{M}_{\text{phy}}^{\prime}$ and $\partial \mathcal{M}_{\text{gs}}^{\prime}$ match entirely (only true for half M\"obius quench, see SM \cite{suppl} Sec.\ VI), quench dynamics will be absent due to the $\tau$ translation symmetry.

\emph{Generic boundary effect.} For generic functions $f(x)$, the boundaries $\partial \mathcal{M}_{\text{phy}}^{\prime}$ and $\partial \mathcal{M}_{\text{gs}}^{\prime}$ may mismatch almost everywhere as shown in \cref{fig1}(b), so the post-quench correlation function cannot be obtained by conformal mapping from correlation in a simple strip geometry $\mathcal{M}_\text{gs}$, which we say has \emph{generic boundary effect} (which includes simple boundary effect as special cases). We circumvent this problem as follows. 

\begin{figure}[tbp]
\begin{center}
\includegraphics[width=3.4in]{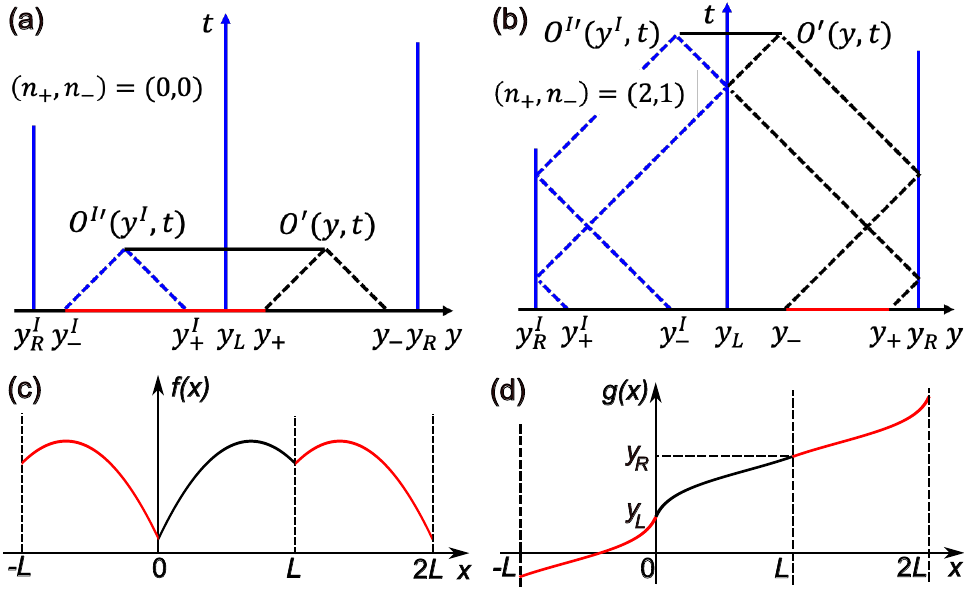}
\end{center}
\caption{(a)-(b): The ancillary mirror PBC system for 
the original system with boundaries
in $y$ coordinates 
for the examples of $(n_+,n_-)=(0,0)$ and $(2,1)$. Index $I$ stands for mirror image. (c)-(d) The equivalent mirror extensions of $f(x)$ and $g(x)$ to $x\in\mathbb{R}$. 
}
\label{fig2}
\end{figure}

We first define an ancillary mirror PBC (MP) quench system with period $2L$ in $x$, where intervals $[0,L]$ and $[-L,0]$ are 
our original quench system
and its mirror copy, respectively (\cref{fig2}(a)-(b)). This allows us to rewrite 
the post-quench $k$-point function in the original geometry with boundaries 
at real time $t\ge0$ as a $2k$-point function of the MP system: $\langle\prod_{j=1}^k\mathcal{O}_j^{\prime}(y_j,t)\rangle=\sqrt{\langle\prod_{j=1}^k\mathcal{O}_j^{\prime}(y_j,t) \mathcal{O}_j^{I\prime}(y_j^I,t)\rangle_\text{mp}}$, where $\mathcal{O}_j^{I\prime}$ is the image operator at mirror position $y_j^I$ of the operator $\mathcal{O}_j^{\prime}$ at position $y_j$. Then, this MP $2k$-point function can be traced back to initial time $t=0$ and calculated from the pre-quench MP correlation via a conformal transformation. 

We sketch below this calculation for 
one-point function
$\langle\mathcal{O}^{\prime}(y,t)\rangle=\sqrt{\langle\mathcal{O}^{\prime}(y,t) \mathcal{O}^{I\prime}(y^I,t)\rangle_\text{mp}}$ with scaling dimension $\Delta$ (see details in SM \cite{suppl} Sec.\ III). In CFT, $\mathcal{O}^{\prime}(y,t)=\mathcal{O}_r^{\prime}(y-t)\overline{\mathcal{O}}_l^{\prime}(y+t)$ decomposes into right-/left-moving parts $\mathcal{O}_r^{\prime}$ and $\overline{\mathcal{O}}_l^{\prime}$. The mirror symmetry ensures equal contributions from the right-/left-moving MP correlations, thus $\langle\mathcal{O}^{\prime}(y,t)\rangle=\langle\mathcal{O}_r^{\prime}(y_2) \mathcal{O}_r^{I\prime}(y_1)\rangle_\text{mp}$ becomes a right-moving MP two-point function, where $y_1=y^I-t$ and $y_2=y-t$ are initial positions of right-moving quasiparticles reaching $(y^I,t)$ and $(y,t)$ in the MP system. As long as $y_1$, $y_2$ are away from $y_L$, $y_R$, they can be locally conformal mapped to the original coordinates $x_1$, $x_2$. Thus, $\langle\mathcal{O}_r^{\prime}(y_2) \mathcal{O}_r^{I\prime}(y_1)\rangle_\text{mp}=\left|\frac{\partial y_1}{\partial x_1}\frac{\partial y_2}{\partial x_2} \right|^{-\frac{\Delta}{2}}\langle\mathcal{O}_r(x_2) \mathcal{O}_r^{I}(x_1)\rangle_\text{mp}$ is the conformal transformation of the initial pre-quench MP correlation, which is $\langle\mathcal{O}_r(x_2) \mathcal{O}_r^{I}(x_1)\rangle_\text{mp}= \left[\frac{\pi \tilde{\epsilon}(x)}{2 L \sin (\pi x_{12}/ 2L)}\right]^{\Delta}$, where $x_{12}=|x_2-x_1|\ (\text{mod }2L)$, and $\tilde{\epsilon}(x)=\tilde{\epsilon}(-x)$ is the effective ultraviolet (UV) cutoff to be determined, which can be $x=g^{-1}(y)$ dependent.

To map the result in terms of $x_{12}$ back to 
the original geometry,
we define the 
light cone coordinates
$x_\pm$ restricted within $[0,L]$ as initial positions of a quasiparticle reaching point $(x,t)$ from the left/right, after possible boundary reflections (note the difference from $x_\pm$ defined below \cref{Ox-simple}). By rewriting $g(x)\mp t-y_L=\mp n_{\pm}(y_R-y_L)+q_{\pm}$ with $n_\pm\in\mathbb{Z}$ and $q_\pm\in[0,y_R-y_L)$, we explicitly have

\begin{equation}\label{reflected w} 
x_\pm = g^{-1}(y_\pm)\ ,\quad y_{\pm}=
\begin{cases}
&y_L+q_{\pm} \  \left(n_{\pm} \text { even}\right) \\ &y_R-q_{\pm} \  \left(n_{\pm} \text { odd}\right)
\end{cases},
\end{equation}
where $n_{\pm}$ is the number of reflections of the quasiparticle by physical boundary. It can then be shown that $x_{12}=\left|x_++(-1)^{n_s}x_-\right|$ or $2L-\left|x_++(-1)^{n_s}x_-\right|$, where $n_s=n_--n_+$ (see examples of \cref{fig2}(a)-(b)). For fixed $t$, $\frac{\partial y_1}{\partial x_1}=\frac{\partial y_-}{\partial x_-}$ and $\frac{\partial y_2}{\partial x_2}=\frac{\partial y_+}{\partial x_+}$. Moreover, $\langle\mathcal{O}^{\prime}(y,0)\rangle$ at $t=0$ should match the initial one-point function of 
the ground state on the finite interval,
which fixes $\tilde{\epsilon}(x)=\epsilon \frac{\partial y}{\partial x}$ for some constant UV cutoff $\epsilon$. This yields a final expression:
\begin{equation}\label{Ox-generic}
\langle\mathcal{O}^{\prime}(y,t)\rangle=\left[ \left|\frac{\partial x_+}{\partial x}\frac{\partial x_-}{\partial x}\right|^{-\frac{1}{2}}\frac{2 L}{\pi\epsilon} \sin \frac{\pi \left|x_++(-1)^{n_s}x_-\right|}{2L}\right]^{-\Delta}
\end{equation}
with $n_s=n_--n_+$ and $x_\pm$ defined in \cref{reflected w}. This is our central result for the generic boundary effect.

The ancillary mirror PBC extension for calculating the correlation function is equivalent to the mirror extension of $f(x)$ from $x\in[0,L]$ into $x\in\mathbb{R}$ as an even function with period $2L$, and accordingly $g(x)$ into $x\in\mathbb{R}$ by \cref{y(x)}, as illustrated in \cref{fig2}(c)-(d). Such an extension of $f(x)$ and $g(x)$ is clearly not an analytical continuation. This indicates the condition for the simple boundary effect (i.e., for \cref{Ox-generic} to reduce to \cref{Ox-simple}) is that the analytical continuation of $f(x)$ in $x\in\mathbb{R}$ is even and $2L$ periodic. This condition is equivalent to our earlier condition for the simple boundary effect that the two boundaries $\partial \mathcal{M}_{\text{phy}}^{\prime}$ and $\partial \mathcal{M}_{\text{gs}}^{\prime}$ match within a finite Euclidean time interval $(-\tau_0,\tau_0)$ (see proof in SM \cite{suppl} Sec.\ IV). This fully clarifies the simple boundary effect from both Euclidean and real time perspectives, and provides an easy criterion for simple or generic boundary effect.

\emph{Entanglement entropy.} As an example, we apply our method to calculate the entanglement entropy $S_A(x,t)=-\text{tr}\left[\rho_A(t)\ln\rho_A(t)\right]$ of a subsystem $A$ defined as the interval $[0,x]$ ($0<x<L$) at post-quench time $t\ge 0$ in our 
system,
where $\rho_A(t)$ is the reduced density matrix. Using the replica trick and the twist operator formalism, $S_A(x,t)$ can be obtained from a one-point function as $S_A(x,t) =-\lim _{n \rightarrow 1} \frac{\partial}{\partial n}\left\langle\mathcal{T}_n^{\prime}(y,t)\right\rangle$ \cite{Calabrese_2004,Cardy_2007, Calabrese_2009b, Castro_Alvaredo_2019, del_Vecchio_del_Vecchio_2024}, 
where $y=g(x)$, and ${\cal T}'_n$ in $w$ coordinate is the twist operator with scaling dimension $\Delta_n = \frac{c}{12} (n-\frac{1}{n})$ in terms of the central charge $c$ and replica index $n$. From \cref{Ox-generic}, we find the post-quench entanglement entropy for generic boundary effect (which reduces to simple boundary effect when the corresponding condition is satisfied):
\begin{align}\label{entanglement entropy final}
\begin{split}
&S_A(x,t)=\frac{c}{6} \ln \left[  \left|\frac{\partial x_+}{\partial x}\frac{\partial x_-}{\partial x}\right|^{-\frac{1}{2}}\frac{2 L}{\pi\epsilon} \sin \frac{\pi \left|x_{+}+(-1)^{n_s}x_{-}\right| }{2L}\right].
\end{split}
\end{align}

\begin{table}[tbp]
\centering
\begin{ruledtabular}
\begin{tabular}{c|c|c}
Quench & $f(x)$ for $x\in[0,L]$  & $g(x)$ for $x\in[0,L]$ \\
\hline
tEH &$\frac{\left(x+L_1\right)\left(L+L_2-x\right)}{L+L_1+L_2}$&$\ln \left(\frac{x+L_1}{L+L_2-x}\right)$\\
tSRD&$\sqrt{\left(x+L_1\right)\left(L+L_2-x\right)}$&$\cos^{-1}{\left(\frac{L+L_2-L_1-2x}{L+L_1+L_2} \right)}$\\
Rainbow&$e^{-k x}$&$\frac{e^{k x}}{k}$\\
M\"obius & $1-\lambda \cos(\frac{2\pi x}{L})$ &$-\frac{L_{\text {eff }}}{\pi} \tan ^{-1}\left(\frac{a}{\tan \frac{\pi x}{L}}\right)$\\
\end{tabular}
\end{ruledtabular}
\caption{Quench functions considered in this work.
For M\"obius quench, $L_{\text{eff}}=
{L}/{\sqrt{1-\lambda^2}}$, 
$a = \sqrt{(1-\lambda)/(1+\lambda)}$,
and taking the limit $\lambda\rightarrow 1$ gives the SSD quench.
}
\label{tbl:fx}
\end{table}

\emph{Examples and numerical verification.} While \cref{entanglement entropy final} is valid for any interacting CFT, we demonstrate its validity by numerically calculating the entanglement entropy of 
a free fermion tight-binding model on an open interval
with quench function $f(x)$ \cite{Peschel_2003}. The
Hamiltonian 
is $H_0=-\frac{1}{2}\sum_{j=1}^{L-1}
(c_{j+1}^{\dagger} c_j+h.c.)$ before quench, and $H=-\frac{1}{2}\sum_{j=1}^{L-1}[f(j)c_{j+1}^{\dagger} c_j+h.c.]$ after quench, where $c_j,c_j^\dag$ are the fermion annihilation/creation operators at $j$th site. We fix the filling at $\nu={1}/{2}$, such that its low energy theory is a CFT with the speed of light $1$ before quench, and central charge $c=1$.
As specific examples of the post-quench Hamiltonian, we consider
the truncated entanglement Hamiltonian (tEH), truncated SRD (tSRD), rainbow and M\"obius (which includes SSD as a limit) quenches \cite{Rodr_guez_Laguna_2017, Wen_2016, MacCormack_2019,Ram_rez_2015}, for which $f(x)$ and $g(x)$ are listed in \cref{tbl:fx} (See SM \cite{suppl} Sec.\ VI). When $L_1=L_2=0$, tEH and tSRD reduce to the entanglement Hamiltonian (EH) and SRD quenches in literature \cite{Wen_2016}. 

\begin{figure}[tbp]
\begin{center}
\includegraphics[width=3.4in]{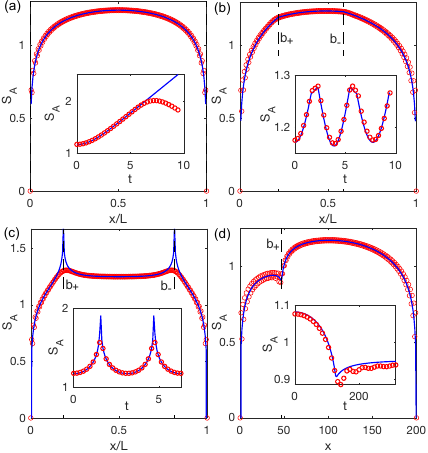}
\end{center}
\caption{
Comparison between the CFT result \cref{entanglement entropy final} with a fixed fitted $\epsilon$ (blue lines) and 
the free-fermion tight-binding calculation (red circles), all for system size $L=200$. Each panel shows $S_A(x,t_0)$ at a fixed time $t=t_0$, and the inset shows $S_A(x_0,t)$ at a fixed position $x=x_0$. The quenches and parameters are: (a) EH (tEH with $L_1=L_2=0$), $t_0=\frac{9\pi}{20}$, $x_0=0.5L$. (b) tEH with $L_1=0.1L$, $L_2=0.3L$, $t_0=\frac{9\pi}{20}$, $x_0=0.5L$. (c) SRD (tSRD with $L_1=L_2=0$), $t_0=\frac{3\pi}{10}$, $x_0=0.5L$. (d) Rainbow (finite $L$) with $k=0.05$, $t_0=60\pi$, $x_0=40$. More examples (including M\"obius and SSD) are shown in SM \cite{suppl} Fig.\ S2.
}
\label{fig3}
\end{figure}

As shown in \cref{fig3} and SM \cite{suppl} Fig.\ S2, $S_A(x,t)$ 
from the CFT formula \cref{entanglement entropy final} (blue lines) and from the tight-binding calculations (red circles) match well, except that the tight-binding data show an additional 
filling-dependent oscillation in $x$ (which is UV physics).

The EH quench shown in \cref{fig3}(a) has simple boundary effect since $y_L=-\infty$ and $y_R=\infty$. Its $S_A(x,t)$ from the CFT \cref{entanglement entropy final} is in a heating phase with perpetual linear growth in $t$, because there exist hot spots $x=0$ and $x=L$ \cite{Fan_2020,Lapierre_2021,Casini_2016}. A hot spot $x_h\in[0,L]$ is where $f(x)\propto|x-x_h|^{\eta}$ with $\eta\ge1$, thus the time $\int^{x_h}\frac{dx}{f(x)}$ for particles to reach $x_h$ diverges, and heat (entropy) is trapped at $x_h$. $S_A(x,t)$ from tight-binding calculations is eventually upper bounded at large $t$, as the spatial UV cutoff (lattice constant) around the hot spots prevents the time divergence.

Figs.\ \ref{fig3}(b)-(d) show the tEH ($L_1,L_2>0$), SRD and rainbow quenches which have generic boundary effect. They have $\partial_x S_A(x,t)$ discontinuities (from the Jacobian
$\frac{\partial x_{\pm}}{\partial x}$ in \cref{entanglement entropy final}) at the vertical dashed line positions $b_+=g^{-1}(y_L+t)$ and $b_-=g^{-1}(y_R-t)$ (recall $y_L=g(0)$, $y_R=g(L)$). This is because $x=b_\pm$ is where $x_\pm$ hits the physical boundary, and $n_\pm$ changes by $1$. Therefore, the discontinuities at $b_\pm$ can be viewed as the shock fronts from boundary reflections. For SRD (\cref{fig3}(c)), the CFT formula shows divergent discontinuities. While tight-binding results cannot diverge due to finite Hilbert space, they approach the CFT formula as $L\rightarrow\infty$ (SM \cite{suppl} Sec. VII). For simple boundary effect (EH in \cref{fig3}(a), M\"obius and SSD in SM \cite{suppl} Fig.\ S2), $S_A(x,t)$ is analytical without discontinuities at any $x\in(0,L)$.

\emph{Discussion.} Our method solves both simple and generic boundary effects of CFT quenches with boundaries,
which correspond to simple and complicated Euclidean path-integral spacetime geometries, respectively. The ancillary mirror PBC picture we developed implies that, the previous solution for CFT quenches with PBC \cite{Moosavi_2021} also corresponds to simple Euclidean path-integral spacetime geometries, which we prove in SM \cite{suppl} Sec.\ V. 
Besides entanglement entropy, \cref{Ox-generic} can be used to calculate any local quantities for generic 
quenches.
$k$-point functions are also in-principle calculable but just complicated \cite{Calabrese_tonni_2009}.

While we have demonstrated our analytical framework through comparisons 
with free-fermion numerics, it also applies to a broad class of interacting conformal field theories, where numerical simulations remain challenging due to rapid entanglement growth during time evolution (the so-called “entanglement barrier”). Notable examples include quantum-critical spin chains and Hubbard-like models. In contrast, our CFT-based approach offers analytical control and insight in such settings. Furthermore, our method enables the study of generic quench problems in quantum simulators—such as Rydberg atom arrays 
\cite{Fendley_2004,
PhysRevA.86.041601,
bernien2017, Keesling_2019,
rader2019floating,
Slagle_2021},
which can be tuned to conformal critical points, providing a promising platform to test our predictions experimentally. 
Looking ahead, 
an intriguing future question is to extend our method to generic time-dependent problems, such as moving mirror and Floquet dynamics problems \cite{
Lapierre_2020a, Lapierre_2020b,
Lapierre_2021,Fan_2020,Akal_2022,wen2018floquet,Martin_2019}. 
Finally, it is interesting to apply our method in the context of holographic duality.
For the case of simple boundary effect, 
Ref.\
\cite{kudlerflam2023bridgingquantumquenchproblems}
constructed 
the holographic dual of 
the M\"obius quench
which 
involves
non-trivial dynamics of
an "end-of-the-world brane",
an extended defect-like object in gravity.
Constructing the holographic dual having 
generic boundary effects
is an intriguing future problem.

\begin{acknowledgments}
\emph{Acknowledgments.} We thank Ruihua Fan, Yingfei Gu and Per Moosavi for helpful discussions.  
B.L. is supported by the Alfred P. Sloan Foundation, the National Science Foundation through Princeton University’s Materials Research Science and Engineering Center DMR-2011750, and the National Science Foundation under award DMR-2141966. S.R.~is supported by the National Science Foundation under 
Award No.\ DMR-2409412.
T.N. is supported by MEXT KAKENHI Grant-in-Aid for Transformative Research Areas A ``Extreme Universe'' (22H05248) and JSPS KAKENHI Grant-in-Aid for Early-Career Scientists (23K13094).
\end{acknowledgments}

\bibliography{cft_ref}

\clearpage
\appendix
\onecolumngrid
\section*{Supplementary Material for ``Inhomogeneous Quantum Quenches of Conformal Field Theory with 
Bondaries"}

\setcounter{equation}{0}
\setcounter{figure}{0}
\setcounter{table}{0}
\makeatletter
\renewcommand{\theequation}{S\arabic{equation}}
\renewcommand{\thefigure}{S\arabic{figure}}

\subsection{I. The twist operator one-point function in the Euclidean path integral picture}

We assume the ground state $|\psi_0\rangle$ of the pre-quench Hamiltonian $H_0$ is non-degenerate. By expressing the initial state as $|\psi_{0}\rangle=\lim _{\beta \rightarrow \infty} e^{-\beta H_0}|\alpha\rangle$ with arbitrary state $|\alpha\rangle$ which has nonzero overlap with the ground state, we can rewrite it in the pre-quench coordinate $z=\tilde{x}+i\tilde{\tau}$ as a path integral in a half-infinite strip $\mathcal{M}_{\text{gs}}^<$, defined as the $\widetilde{\tau}<0$ region of strip $\mathcal{M}_{\text{gs}}$ in the main text Fig. 1(a) 
with straight boundaries $\partial\mathcal{M}_{\text{gs}}$ at $\tilde{x}=0$ and $L$: 

\begin{equation}\label{ground state*}
\langle \phi_F|\psi_0\rangle=\int_{\phi_{\tilde{\tau}=0}=\phi_F} D \phi e^{-\int_{-\infty}^0 d \tilde{\tau} \int_0^L d\tilde{x} \mathcal{L}_0(\phi)}\ ,
\end{equation}
where $\phi(\tilde{x},\tilde{\tau})$ represents all the quantum fields in the CFT. The uniform Lagrangian $\mathcal{L}_0(\phi (\tilde{x},\tilde{\tau}))$ lives in the strip $\tilde{x} \in[0, L]$, and is the Legendre transformation of the uniform energy density $h(\tilde{x})$. The state $|\phi_F\rangle$ appearing in the path integral is defined as the coherent state of the quantum fields $\phi$.

Since the theory we consider is CFT, the Lagrangian density $\mathcal{L}_0(\phi)$ has scaling dimension 2. Therefore, under conformal mapping $z \rightarrow w=g(z)$, with  $dw/dz=dg(z)/dz=1/f(z)$:
\begin{align}
\begin{split}
\mathcal{L}_0(\phi(z)) \longrightarrow \frac{1}{|f(z)|^2} \mathcal{L}_0^{\prime}\left(\phi^{\prime}(w)\right)\ ,\qquad 
d \tilde{\tau} d\tilde{ x}=|d z|^2=|f(z)|^2|d w|^2=|f(z)|^2 d \tau d y\ ,
\end{split}
\end{align}
so the action of the path integral \cref{ground state*} transforms as 
\begin{align}
\begin{split}
&\int_{-\infty}^0 d \tilde{\tau} \int_0^L d \tilde{x} \mathcal{L}_0(\phi)=\int_{ \mathcal{M}_\text{gs}^{\prime <}} d \tau d y |f(z)|^2 \frac{1}{|f(z)|^2} \mathcal{L}_0^{\prime}\left(\phi^{\prime}\right)=\int_{ \mathcal{M}_\text{gs}^{\prime <}} d \tau d y \mathcal{L}_0^{\prime}\left(\phi^{\prime}\right)
\end{split}
\end{align}
where the new Lagrangian $\mathcal{L}_0^{\prime}\left(\phi^{\prime}\right)$ is still uniform in spacetime due to conformal symmetry, and $\phi'(w)$ represents quantum fields in the $w$ coordinate. Here $\mathcal{M}_\text{gs}^{\prime <}$ denotes the $\tau<0$ region of the manifold $\mathcal{M}_\text{gs}^{\prime}$ in main text Fig. 1(b), and it has curved boundaries $\partial \mathcal{M}_\text{gs}^{\prime}$ mapped from the straight boundaries $\partial \mathcal{M}_\text{gs}$ in main text Fig. 1(a). The path integral for the initial state now becomes
\begin{align}\label{initial state w*}
\begin{split}
\langle \phi_F'|\psi_0\rangle=\int_{\phi_{\tau=0}'=\phi_F'} D \phi^{\prime} e^{-\int_{ \mathcal{M}_{\text{gs}}^{\prime <}} d \tau d y \mathcal{L}_0^{\prime}\left(\phi^{\prime}\right)}\ ,
\end{split}
\end{align}

Meanwhile, the conformal mapping $z \rightarrow w=g(z)$ also maps the energy density (which has scaling dimension $2$) as
\begin{equation}\label{seq-hw}
h(z)=\frac{1}{|f(z)|^2}h^{\prime}(w)   \ ,
\end{equation}
where $h'(w)$ is the uniform energy density in the $w=y+i\tau$ coordinates. In particular, at time $\tau=it=0$, one has $w=y$, and the corresponding $z=\tilde{x}=x$ (recall that coordinates $\tilde{x}$ and $x$ match at time $t=0$), so \cref{seq-hw} implies
\begin{equation}\label{seq-hy}
h(x)=\frac{1}{|f(x)|^2}h^{\prime}(y)\ .
\end{equation}
Therefore, the post-quench time evolution operator $e^{-H\tau}$ of $|\psi_0\rangle$ can be transformed into the $(\tau,y)$ coordinates as
\begin{align}\label{time evolution*}
\begin{split}
e^{- H \tau}&=e^{-\tau\int_0^L f(x) h(x) d x }=e^{-\tau\int_{y_{L}}^{y_{R}} h^{\prime}(y)dy }\ ,
\end{split}
\end{align}
where we have used \cref{seq-hy} and the fact that $dx=f(x)dy$. This is equivalent to a path integral in coordinate $w=y+i\tau$ with straight boundaries 
\begin{align}
\begin{split}\label{time evolution w*}
\langle \phi_F'|e^{-H\tau}| \phi_I'\rangle&=\int_{\phi_{\tau=0}'=\phi_I'}^{\phi_{\tau}'=\phi_F'} D \phi^{\prime}e^{-\int_{0}^{\tau}d \tau' \int_{y_{L}}^{y_{R}} d y \mathcal{L}_0^{\prime}\left(\phi^{\prime}\right)}
\end{split}
\end{align}
Therefore, the state at time $t=-i\tau$ is equivalent to time evolution path integral in the $w=(\tau,y)$ coordinates with uniform Lagrangian $\mathcal{L}_0^{\prime}$ and straight boundary $\partial\mathcal{M}'_\text{phy}$, from a ground state $|\psi_0\rangle$ represented by a path integral with curved boundary $\partial\mathcal{M}'_\text{gs}$. This yields a path integral representation of the k-point function $\left\langle\mathcal{O}_j^{\prime}(w_j)\right\rangle$ with $w_j=y_j+i\tau$:
\begin{equation}
\langle\prod_{j=1}^k\mathcal{O}_j^{\prime}(w_j)\rangle_{\mathcal{M}_\text{gs}'}=\int D \phi^{\prime} e^{-\int_{ \mathcal{M}_{\text{gs}}^{\prime <}} d \tau d y \mathcal{L}_0^{\prime}\left(\phi^{\prime}\right)} e^{-\int_{ \mathcal{M}_{\text{gs}}^{\prime >}} d \tau d y \mathcal{L}_0^{\prime}\left(\phi^{\prime}\right)} e^{\int_{0}^{\tau}d \tau' \int_{y_{L}}^{y_{R}} d y \mathcal{L}_0^{\prime}\left(\phi^{\prime}\right)} \prod_{j=1}^k\mathcal{O}_j^{\prime}(y_j) e^{-\int_{0}^{\tau}d \tau' \int_{y_{L}}^{y_{R}} d y \mathcal{L}_0^{\prime}\left(\phi^{\prime}\right)}
\end{equation}
where $\mathcal{M}_\text{gs}^{\prime >}$ denotes the $\tau>0$ region of the manifold $\mathcal{M}_\text{gs}^{\prime}$ (the manifold for path integral representation of the bra state $\langle \psi_0|$). This is as represented in the main text Fig. 1(b) in the path integral order of $1\rightarrow 2\rightarrow 3\rightarrow 4$.

\subsection{II. Derivation of entanglement entropy with the simple boundary effect}
For simple boundary effect (the case the two boundaries $\partial \mathcal{M}_{\text{phy}}^{\prime}$ and $\partial \mathcal{M}_{\text{gs}}^{\prime}$ match within a finite Euclidean time interval $(-\tau_0,\tau_0)$ with $\tau_0>0$), we showed in the main text that, the $k$-point function $\langle\prod_{j=1}^k\mathcal{O}_j^{\prime}(w_j)\rangle_{\mathcal{M}_\text{gs}'}$ is calculable from the $k$-point function  $\langle\prod_{j=1}^k\mathcal{O}_j(z_j)\rangle_{\mathcal{M}_\text{gs}}$ of Euclidean path integral in the strip $\mathcal{M}_\text{gs}$ (defined as the manifold $\text{Re}(z)\in[0,L]$) in main text Fig. 1. This $k$-point function can be derived by a further conformal mapping from the strip into the half plane, which gives (for simplicity we assume all operators have zero conformal spin, e.g., nonchiral)
\begin{align}\label{k-point function}
\langle\prod_{j=1}^k\mathcal{O}_j^{\prime}(w_j)\rangle_{\mathcal{M}_\text{gs}'}=\prod_{j=1}^{k}\left|\frac{dw_j}{dz_j}\right|^{-\Delta_j} \langle\prod_{j=1}^k\mathcal{O}_j(z_j)\rangle_{\mathcal{M}_\text{gs}}
\end{align}
where $z_j=g^{-1}(w_j)$, and $\Delta_j$ is the scaling dimension of $\mathcal{O}_j^{\prime}$ which maps to $\mathcal{O}_j$ in $z$ coordinate.

In particular, from \cite{Calabrese_2009b}, one-point function of an operator $\mathcal{O}$ with scaling dimension $\Delta$ in a strip in the $z$ coordinate is given by
\begin{align}\label{one-point operator in z*}
\mathcal{O}(z)=\left[\frac{\pi \tilde{\epsilon}(z)}{2 L \sin (\pi \tilde{x}/ L)}\right]^{\Delta}\ , \qquad \tilde{x}=\text{Re}(z)\ ,
\end{align}
where $\tilde{\epsilon}(z)$ is the UV cutoff which could be $z$-dependent in our inhomogeneous CFT problem. We will determine $\tilde{\epsilon}(z)$ later. The one point function in the $w=g(z)$ coordinate is then given by
\begin{equation}\label{one-point operator in w*}
\left\langle\mathcal{O}^{\prime}(w)\right\rangle=\left|\frac{d w}{d z}\right|^{-\Delta}\left\langle\mathcal{O}(z)\right\rangle_{\mathcal{M}_\text{gs}}=
\left[ \left|\frac{d w}{d z}\right|^{-1} \frac{\pi \tilde{\epsilon}(z)}{2 L \sin (\pi \tilde{x}/ L)}\right]^{\Delta}\ .
\end{equation}

Here we have
\begin{equation}
y=g(x)\ , \qquad w=y+i\tau\ ,\qquad \tilde{x}(y,\tau)=\text{Re}(z)=\text{Re}\left[g^{-1}(y+i\tau)\right]\ ,\qquad \frac{dw}{dz}=\frac{1}{f(z)}=\frac{1}{f(g^{-1}(y+i\tau))}\ .
\end{equation}
Note that both $
\tilde{x}$ and the Jacobian $d w/ d z$ can be time $\tau$ dependent. 
The UV cutoff $\tilde{\epsilon}(z)$ in \cref{one-point operator in w*} is fixed by requiring that at $\tau=t=0$, the expression should match with that of the ground state of $H_{0}$ on 
the spatial interval $[0,L]$,
which is a known result from literature \cite{Calabrese_2009b}. This fixes the UV cutoff as
\begin{align}\label{uv cutoff*}
\tilde{\epsilon}(z)= \epsilon|d w / d z|_{\tau=0}\ ,
\end{align}
where $\epsilon$ is a fixed constant. The one point function in the $w=g(z)$ coordinate in \cref{one-point operator in w*} is now given by
\begin{equation}\label{UV-fixed one-point operator in w*}
\left\langle\mathcal{O}^{\prime}(w)\right\rangle=\left|\frac{d w}{d z}\right|^{-\Delta}\left\langle\mathcal{O}(z)\right\rangle_{\mathcal{M}_\text{gs}}=
\left[\frac{|d w / d z|_{\tau}}{|d w / d z|_{\tau=0}} \frac{2 L \sin (\pi  \tilde{x}/ L)}{\pi \epsilon}\right]^{-\Delta}\ .
\end{equation}

We can analytical continue the above result into the real coordinates \cite{Lapierre_2021}. For this purpose, we assume $g^{-1}(w)$ is analytical on the real axis, which can be proved in Sec. IV 
under the condition for simple boundary effect. In this case, $w=y+i\tau$ and $\bar{w}=y-i\tau$ become light cone coordinates $y_\pm=y\mp t$, while $z=g^{-1}(w)$ and $\bar{z}=g^{-1}(\bar{w})$ becomes
light cone coordinates $x_\pm = g^{-1}(y_\pm)=g^{-1}(y\mp t)= g^{-1}(g(x) \mp t)$. Now we have for fixed $\tau=it$,
\begin{equation}
\left|\frac{d w}{ d z}\right|_{\tau}=\sqrt{\frac{dw}{dz}\frac{d\Bar{w}}{d\Bar{z}}}\Big|_{t}=\sqrt{\frac{dy_{+}}{dx_{+}}\frac{dy_{-}}{dx_{-}}}\Big|_{t}\ ,\qquad \left|\frac{d w}{ d z}\right|_{\tau=0}=\frac{dy}{dx}\ .
\end{equation}
Since $dy_{+}=dy=dy_{-}$ for fixed $t=-i\tau$, we have
\begin{equation}
\frac{|d w / d z|_{\tau}}{|d w / d z|_{\tau=0}}=\left(\frac{\partial x_+}{\partial x}\frac{\partial x_-}{\partial x}\right)^{-\frac{1}{2}}\ , \qquad \tilde{x}=\frac{z+\Bar{z}}{2}=\frac{x_++x_-}{2}\ ,
\end{equation}
and thus \cref{UV-fixed one-point operator in w*} becomes
\begin{equation}\label{Ox-simple-S}
\langle\mathcal{O}^{\prime}(y,t)\rangle=\left[ \left|\frac{\partial x_+}{\partial x}\frac{\partial x_-}{\partial x}\right|^{-\frac{1}{2}}\frac{2 L}{\pi\epsilon} \sin \frac{\pi \left(x_++x_-\right)}{2L}\right]^{-\Delta},
\end{equation}

As an example, we apply our method to calculate the twist operator one-point function $\left\langle\mathcal{T}_n^{\prime}(w)\right\rangle$. The twist operator has scaling dimension $\Delta_n= \frac{c}{12}\left(n-\frac{1}{n}\right)$. Explicitly, by taking the $n\rightarrow 1$ limit, we find the entanglement entropy given as follows:
\begin{align}\label{entanglement entropy*}
S_A(x,-i\tau) & =-\lim _{n \rightarrow 1} \frac{\partial}{\partial n}\left\langle\mathcal{T}_n^{\prime}(w)\right\rangle =S_A(x,t) =\frac{c}{6} \ln \left[ \left(\frac{\partial x_+}{\partial x}\frac{\partial x_-}{\partial x}\right)^{-\frac{1}{2}}\frac{2 L}{\pi\epsilon} \sin \frac{\pi \left(x_++x_-\right)}{2L}\right]\ .
\end{align}

\subsection{III. Derivation of entanglement entropy with the generic boundary effect}
In the generic boundary effect case, we have defined the lightcone coordinates taking into account the boundary reflections in main text Eq. (8), which we rewrite here for convenience:
By rewriting $g(x)\mp t-y_L=\mp n_{\pm}(y_R-y_L)+q_{\pm}$ with $n_\pm\in\mathbb{Z}$ and $q_\pm\in[0,y_R-y_L)$, we explicitly have
\begin{equation}\label{reflected-w*} 
\begin{split}
x_\pm &= g^{-1}(y_\pm)\ ,\quad y_{\pm}=
\begin{cases}
&y_L+q_{\pm} \  \left(n_{\pm} \text { even}\right) \\ &y_R-q_{\pm} \  \left(n_{\pm} \text { odd}\right)
\end{cases}, \\ 
y\mp t-y_L &=g(x)\mp t-y_L=\mp n_{\pm}(y_R-y_L)+q_{\pm}\ ,\quad n_\pm\in\mathbb{Z}\ ,\quad q_\pm\in[0,y_R-y_L)\ .
\end{split}
\end{equation}

For generic boundary effect, we showed in the main text (Fig. 2) that the post-quench 
$k$-point function on the interval
at real time $t\ge0$ can be calculated as a $2k$-point function of the ancillary mirror PBC (MP) quench problem:
\begin{equation}
\langle\prod_{j=1}^k\mathcal{O}_j^{\prime}(y_j,t)\rangle=\sqrt{\langle\prod_{j=1}^k\mathcal{O}_j^{\prime}(y_j,t) \mathcal{O}_j^{I\prime}(y_j^I,t)\rangle_\text{mp}}
\end{equation}
where $\mathcal{O}_j^{I\prime}$ is the image operator at mirror position $y_j^I=2y_{L}-y_j$ $(\text{mod }2(y_R-y_L))$ of the operator $\mathcal{O}_j^{\prime}$ at position $y_j$. Compared to the previously studied PBC quench problems \cite{Moosavi_2021, Lapierre_2021} which have smooth deformation functions, our mirror PBC problem by definition has a deformation function $f(x)$ from mirror extension which is not analytical at $x=0$ and $x=L$, as shown in main text Fig. 2(c). However, this does not affect anything as long as $x$ and $x_\pm$ are not at the non-analytical points. 

In particular, for 
a one-point function
of a nonchiral operator, 
\begin{equation}\label{seq-general-onepoint}
\langle\mathcal{O}^{\prime}(y,t)\rangle=\sqrt{\langle\mathcal{O}^{\prime}(y,t) \mathcal{O}^{I\prime}(y^I,t)\rangle_\text{mp}}
\end{equation}

In the $(y,t)$ coordinate, the post-quench energy density is uniform, and the speed of light is $1$ everywhere, so the two-point function is simple. If we identify $y$ with $y+2(y_R-y_L)$ imposing the $2(y_R-y_L)$ spatial period in $y$, the original nonchiral operator can be decomposed as $\mathcal{O}^{\prime}(y,t)=\mathcal{O}_r^{\prime}(y-t)\overline{\mathcal{O}}_l^{\prime}(y+t)$, where the right-/left-moving parts are $\mathcal{O}_r^{\prime}$ and $\overline{\mathcal{O}}_l^{\prime}$ respectively. For nonchiral operators, the operator $\mathcal{O}^{\prime}$ has left/right scaling dimensions $(\frac{\Delta}{2},\frac{\Delta}{2})$, so both $\mathcal{O}_r^{\prime}$ and $\mathcal{O}_l^{\prime}$ have chiral scaling dimension $\Delta/2$. Similar decomposition holds for the image operator $\mathcal{O}^{I\prime}$. In Euclidean time, this is equivalent to decomposition of the two-point function into holomorphic and anti-holomorphic parts. Accordingly, the two-point function can be decomposed as
\begin{equation}
\langle\mathcal{O}^{\prime}(y,t)\mathcal{O}^{I\prime}(y^I,t)\rangle_{\text{mp}}= \langle\mathcal{O}_{r}^{\prime}(y-t)\mathcal{O}_{r}^{I\prime}(y^I-t)\rangle_{\text{mp}} \langle\overline{\mathcal{O}}_{l}^{\prime}(y+t)\overline{\mathcal{O}}_{l}^{I\prime}(y^I+t)\rangle_{\text{mp}}\ ,
\end{equation}
where the left-moving part $\langle\mathcal{O}_{r}^{\prime}(y-t)\mathcal{O}_{r}^{I\prime}(y^I-t)\rangle_{\text{mp}}$ and right-moving part $\langle\overline{\mathcal{O}}_{l}^{\prime}(y+t)\overline{\mathcal{O}}_{l}^{I\prime}(y^I+t)\rangle_{\text{mp}}$ are expectation values from the initial state. By mirror symmetry of the ancillary mirror PBC system, $\langle\mathcal{O}_{r}^{\prime}(y-t)\mathcal{O}_{r}^{I\prime}(y^I-t)\rangle_{\text{mp}}=\langle\overline{\mathcal{O}}_{l}^{\prime}(y+t)\overline{\mathcal{O}}_{l}^{I\prime}(y^I+t)\rangle_{\text{mp}}$, and thus \cref{seq-general-onepoint} 
becomes
\begin{equation}
\langle\mathcal{O}^{\prime}(y,t)\rangle=\sqrt{\langle\mathcal{O}^{\prime}(y,t)\mathcal{O}^{I\prime}(y^I,t)\rangle_{\text{mp}}}= \langle\mathcal{O}_{r}^{\prime}(y-t)\mathcal{O}_{r}^{I\prime}(y^I-t)\rangle_{\text{mp}}\ .
\end{equation}

At time $t=0$, the two points $y_1=y^I-t$ $(\text{mod }2(y_R-y_L))$ and $y_2=y-t$ $(\text{mod }2(y_R-y_L))$ are the two end-points of the red interval shown in main text Fig. 2(a)-(b), which are initial positions of right-moving quasiparticles reaching positions $(y^I,t)$ and $(y,t)$, respectively. Depending on the number of reflections $n_\pm$ in \cref{reflected-w*}, the two points can be $y_\pm$ or $y_\pm^I=2y_L-y_\pm$. In the mirror-extended $x$ coordinate, we assume $y_1$ and $y_2$ map to two points $x_1$ and $x_2$, which can be $x_\pm$ or $-x_\pm$ accordingly. More explicitly, one has
\begin{equation}  
y_1=y^I-t=\begin{cases}
&y_-\ (\text{mod }2(y_R-y_L))\ ,\quad (n_-\text{ odd}) \\
&y_-^I\ (\text{mod }2(y_R-y_L))\ ,\quad (n_-\text{ even})
\end{cases},\quad
y_2=y-t=\begin{cases}
&y_+\ (\text{mod }2(y_R-y_L))\ ,\quad (n_+\text{ even}) \\
&y_+^I\ (\text{mod }2(y_R-y_L))\ ,\quad (n_+\text{ odd})
\end{cases},
\end{equation}
and accordingly, their mappings in the mirror-extended $x$ coordinate are
\begin{equation}
x_1=-(-1)^{n_-}x_-\ (\text{mod }2L)\ ,\qquad x_2=(-1)^{n_+}x_+\ (\text{mod }2L)\ .
\end{equation}
The red interval between $y_1$ and $y_2$ in main text Fig. 2(a)-(b), when mapped back to the $x$ coordinates, is the interval between $x_1$ and $x_2$ $(\text{mod }2L)$. Since
\begin{equation}
x_2-x_1=(-1)^{n_+}x_++(-1)^{n_-}x_-\ (\text{mod }2L)\ ,
\end{equation}
with $x_\pm\in[0,L]$, we find the red interval has a length $x_{12}=|x_2-x_1|\ (\text{mod }2L)$ in the $x$ coordinate
\begin{equation}\label{seq-delta12}
x_{12}=\begin{cases}
&\left|x_++(-1)^{n_s}x_-\right| \ ,\qquad\qquad\   (\text{if\ } n_s=1\text{ or }n_-=n_+\in 2\mathbb{Z})\\ 
&2L-\left|x_++(-1)^{n_s}x_-\right|\ ,\qquad (\text{if\ } n_s=-1\text{ or }n_-=n_+\in 2\mathbb{Z}+1)
\end{cases}\ .
\end{equation}
where $n_s=n_--n_+$.

The initial state $|\psi_0\rangle$ of the original 
system
is the uniform ground state of uniform Hamiltonian $H_0$ in the $x$ coordinate. Correspondingly, the initial state (mirror extension of $|\psi_0\rangle$) of the ancillary mirror PBC problem will be the ground state of uniform Hamiltonian (mirror extension of $H_0$) with PBC and spatial period $2L$ in the $x$ coordinate. Therefore, in the $x$ coordinate, the chiral two-point function of $\mathcal{O}_r$ and $\mathcal{O}_r^{I}$ (which have scaling dimension $\Delta/2$) of the initial state is easy to calculate (via a further mapping into two-point function in the whole plane in the Euclidean formalism), which results \cite{Calabrese_2009b}
\begin{equation}
\langle\mathcal{O}_{r}(x_2)\mathcal{O}^{I}_{r}(x_1)\rangle_{\text{mp}}=\left[\frac{\pi \tilde{\epsilon}(x)}{2 L \sin (\pi x_{12}/ 2L)}\right]^{\Delta}\
\end{equation}
where $\tilde{\epsilon}(x)=\tilde{\epsilon}(-x)$ is the effective (mirror symmetric) UV cutoff which can be position $x$ dependent (similar to the case of simple boundary effect), since we are considering an inhomogeneous CFT quench. As a result, the two-point function in the $y$ coordinate is related by conformal transformation:
\begin{equation}
\langle\mathcal{O}_{r}^{\prime}(y_2)\mathcal{O}_{r}^{I\prime}(y_1)\rangle_{\text{mp}}=\left(\frac{\partial y_1}{\partial x_1}\frac{\partial y_2}{\partial x_2}\right)^{-\frac{\Delta}{2}}\langle\mathcal{O}_{r}(x_2)\mathcal{O}_{r}^{I}(x_1)\rangle_{\text{mp}}=\left[ \left(\frac{\partial y_+}{\partial x_+}\frac{\partial y_-}{\partial x_-}\right)^{-\frac{1}{2}}\frac{\pi \tilde{\epsilon}(x)}{2 L \sin (\pi x_{12}/ 2L)}\right]^{\Delta},
\end{equation}
where we have used the fact that $\partial y_1/\partial x_1=\partial y_-/\partial x_-$ and $\partial y_2/\partial x_2=\partial y_+/\partial x_+$, understood as the derivatives at fixed time $t$. The cutoff $\tilde{\epsilon}(x)$ can again be determined by the requirement that at $t=0$, the expression should be equal to that of the initial state, which gives
\begin{equation}
\tilde{\epsilon}(x)=\epsilon\frac{\partial y}{\partial x}=\epsilon\left|\frac{\partial y_+}{\partial x}\frac{\partial y_-}{\partial x}\right|^{\frac{1}{2}}\ , \qquad \rightarrow \qquad \left(\frac{\partial x_+}{\partial y_+}\frac{\partial x_-}{\partial y_-}\right)^{-\frac{1}{2}}\frac{1}{\tilde{\epsilon}(x)}=\left|\frac{\partial x_+}{\partial x}\frac{\partial x_-}{\partial x}\right|^{-\frac{1}{2}}\frac{1}{\epsilon}\ ,
\end{equation}
where $\epsilon$ is a constant cutoff, and we have used the fact that $|\partial y_\pm/\partial y|=1$. Together with the expression of $x_{12}$ in \cref{seq-delta12}, we arrive at the final expression for 
the one-point function:
\begin{equation}\label{Ox-generic-S}
\langle\mathcal{O}^{\prime}(y,t)\rangle=\left[ \left|\frac{\partial x_+}{\partial x}\frac{\partial x_-}{\partial x}\right|^{-\frac{1}{2}}\frac{2 L}{\pi\epsilon} \sin \frac{\pi \left|x_++(-1)^{n_s}x_-\right|}{2L}\right]^{-\Delta}
\end{equation}
with $n_s=n_--n_+$ and $x_\pm$ defined in \cref{reflected-w*}. This is our central result for the generic boundary effect. 

As an example, we apply our method to calculate 
the twist operator one-point function $\langle\mathcal{T}_n^{\prime}(y,t)\rangle$ on the interval, 
which can be further used to calculate the entanglement entropy. The twist operator has scaling dimension $\Delta_n= \frac{c}{12}\left(n-\frac{1}{n}\right)$. Explicitly, by taking the $n\rightarrow 1$ limit, we find the entanglement entropy given as follows:
\begin{align}\label{entanglement entropy-generic*}
S_A(x,t) & =-\lim _{n \rightarrow 1} \frac{\partial}{\partial n}\left\langle\mathcal{T}_n^{\prime}(y,t)\right\rangle =\frac{c}{6} \ln \left[  \left|\frac{\partial x_+}{\partial x}\frac{\partial x_-}{\partial x}\right|^{-\frac{1}{2}}\frac{2 L}{\pi\epsilon} \sin \frac{\pi \left|x_{+}+(-1)^{n_s}x_{-}\right| }{2L}\right].
\end{align}
which is the result we showed in the main text. 

The above derivation, which used analyticity of $x_\pm$ as a function of $x$ and $t$, is valid as long as $x_\pm$ do not hit the boundaries $x=0$ or $x=L$ (where the function is not necessarily analytical). However, the derived entanglement entropy $S_A$ in \cref{entanglement entropy-generic*} is continuous when $x_\pm$ hit the boundaries. $\partial_xS_A$ and $\partial_t S_A$ are often discontinuous when $x_\pm$ hit the boundaries, due to discontinuity of the second derivatives of $x_\pm$ as a function of $x$ and $t$.

\subsection{IV. Two equivalent conditions for the simple boundary effect}\label{sec:Equivalence}

Given $f(x)$ real and analytic on $[0,L]$ (which is the assumption for our 
quench problem
to have simple boundary effect), we here prove the equivalence of the two following conditions for simple boundary effect we used in the main text: 

(1) the two boundaries $\partial \mathcal{M}_{gs}^{\prime}$ and $\partial \mathcal{M}_{phy}^{\prime}$ in the $w=y+i\tau$ (mapped from $w=g(z)$) coordinate match for a finite Eucliean time interval $\tau\in(-\tau_0,\tau_0)$;

(2) The analytic continuation of $f(x)$ on the real axis $x\in\mathbb{R}$ is an even function with a period $2L$, namely, $f(x)=f(-x)=f(x+2L)$.

--- We first prove that condition (2) leads to condition (1). If $f(x)$ is even and have a period of $2L$, then the conformal transformation function $y=g(x)=\int^x \frac{d x^{\prime}}{f\left(x^{\prime}\right)}$ would satisfy 
\begin{equation}\label{seq-g-property}
g(x)-y_L=-[g(-x)-y_L]\ ,\quad g(x+2L)-g(x)=g(2L)-g(0)=2(y_{R}-y_{L})\ ,
\end{equation}
where $y_L=g(0)$ and $y_R=g(L)$. 

Since $f(x)$ is analytical in $x\in\mathbb{R}$ (due to analyticity in $x\in[0,L]$ and the fact it is even and periodic), the complex function $w=g(z)=\int^z \frac{d z^{\prime}}{f(z^{\prime})}$ is holomorphic, and thus $w=g(z)$ has a Talor series expansion around the real axis $\text{Im}z=0$. Around $z=0$, we have
\begin{equation}\label{w expansion around 0}
    w=g(z)= y_L+\sum_{n=1}^\infty a_n z^n
\end{equation}
where $g(0)=y_{L}$, and the coefficients $a_n$ are real because $g(x)$ maps the real axis to the real axis. Meanwhile, \cref{seq-g-property} tells us $g(x)-y_L=-[g(-x)-y_L]$, which implies that $a_n = 0$ if $n$ is even. Therefore, the left boundary of $\partial \mathcal{M}_{gs}$ in the $z$ coordinate, which is the imaginary axis $z=i\tilde{\tau}$ (with $\tilde{\tau}\in\mathbb{R}$), maps to a purely imaginary $w-y_L=g(z)-y_{L}$, indicating that a left boundary $w=y_L+i\tau$ ($\tau\in\mathbb{R}$) of $\partial \mathcal{M}_{gs}^{\prime}$, for small $\tau$ around $0$ where the Taylor expansion \cref{w expansion around 0} converges. This is exactly the same as the left boundary of $\partial \mathcal{M}_{phy}^{\prime}$.

Similarly, around $z=L$, $g(z)$ has the Taylor expansion
\begin{equation}\label{w expansion around L}
    w=g(z)=y_R+\sum_{n=1}^\infty b_n (z-L)^n
\end{equation}
where $g(L)=y_{R}$ and $b_n$ are real. Since $g(x)-y_L=-[g(-x)-y_R]$ and $g(x+2L)-g(x)=2(y_{R}-y_{L})$ as given in \cref{seq-g-property}, one has $g(L+x)-y_{R}=-[g(L-x)-y_{R}]$, and thus $b_n = 0$ if $n$ is even. Therefore, $g(z)-y_{R}$ is purely imaginary for $z=L+i\tilde{\tau}$ ($\tilde{\tau}\in\mathbb{R}$). By the same argument around $z=0$ above, we conclude that the right boundaries of $\partial \mathcal{M}_{gs}^{\prime}$ and $\partial \mathcal{M}_{phy}^{\prime}$ match at small $\tau$ where the Taylor expansion \cref{w expansion around L} converges.

Moreover, since $f(z)$ (and thus $g(z)$) is analytical on the real axis, the boundary $\partial \mathcal{M}_{gs}$ in the $z$ coordinate cannot be mapped onto any point on the real axis interval $w=y\in(y_L,y_R)$ in the $w$ coordinates. Otherwise, there are at least two distinct points in the $z$ coordinates mapping to a single point on the real axis in the $w$ coordinates, contradicting with the analyticity of $g(z)$ on the real axis. Therefore, the mapped boundary $\partial \mathcal{M}_{phy}^{\prime}$ will not touch the real axis $w=y\in(y_L,y_R)$ inside the boundary $y=y_L$ and $y=y_R$.

All together, the above reasoning shows that there is a finite Euclidean time interval $(-\tau_0,\tau_0)$ in which $\partial \mathcal{M}_{gs}^{\prime}$ and $\partial \mathcal{M}_{phy}^{\prime}$ match.

--- Conversely, we can show that condition (1) leads to condition (2). First, $f(x)$ being real and analytic (our assumption) implies Taylor expansions of the same form as \cref{w expansion around 0,w expansion around L}, with real coefficients $a_n$ and $b_n$. If the two boundaries $\partial \mathcal{M}_{gs}^{\prime}$ and $\partial \mathcal{M}_{phy}^{\prime}$ match within a finite $\partial \mathcal{M}_{gs}^{\prime}$ and $\partial \mathcal{M}_{phy}^{\prime}$ match, it indicates that $z=i\tilde{\tau}$ and $z=L+i\tilde{\tau}$ ($\tilde{\tau}\in\mathbb{R}$) map to $w=y_L+i\tau$ and $z=L+i\tau$ ($\tau\in\mathbb{R}$) for small $\tilde{\tau}$, respectively. Therefore, $a_n=0$ and $b_n=0$ if $n$ is even. This then implies the property of $g(x)$ in \cref{seq-g-property} for real $x$. In particular, because $g(x)$ is well-defined in $x\in[0,L]$, the property of \cref{seq-g-property} near $x=0$ and $x=L$ implies an analytical continuation into the entire real axis $x\in\mathbb{R}$. Accordingly, we conclude that $f(x)$, as the inverse derivative of $g(x)$, has an analytical continuation on the entire real axis as an even function with a period $2L$.

We have thus proved the equivalence of conditions $(1)$ and $(2)$.

Simple examples of this class of even analytical functions with a period of $2L$ in $x\in\mathbb{R}$ are:
\begin{equation}\label{even f}
f(x)=\sum_{n=0}^N \lambda_n \cos\left(\frac{n\pi x}{L}\right)\ ,
\end{equation}
where $\lambda_n$ are real coefficients, and $N$ is a positive integer. Both the M\"obius function $f(x)=1-\lambda\cos\left(\frac{2\pi x}{L}\right)$ and the half-M\"obius function $f(x)=1-\lambda\cos\left(\frac{\pi x}{L}\right)$ belong to this class of functions.

\subsection{V. Euclidean path-integral spacetime geometry for smooth inhomogeneous CFT quenches with PBC}

As another note, we show here that for inhomogeneous CFT quenches with PBC of period $L$ which have a non-negative smooth deformation function $f(x)$ in $x\in[0,L]$ (which have been studied before \cite{Moosavi_2021}), their Euclidean spacetime geometry for calculating path integrals is a simple geometry in a finite imaginary time $\tau$ interval $[-\tau_0,\tau_0]$. In this sense, the PBC quench problems with smooth deformations are similar to 
quench problems
with simple boundary effect we identified in this paper.

For the PBC quench problem, we can similarly define the conformal mapping from coordinate $z=\tilde{x}+i\tilde{\tau}$ to $w=y+i\tau=g(z)$, where $g(z)$ is the analytical continuation of $g(x)=\int^x \frac{dx'}{f(x')}$. The initial state, which is the ground state of the uniform CFT Hamiltonian with PBC, can be written as a Euclidean path integral in the cylinder $\mathcal{M}_{\text{gs}}$ with constant circumference $L$ in the $z$ coordinate. Assume it maps into a manifold $\mathcal{M}_{\text{gs}}'$ in the $w$ coordinate. We define $y_L=g(0)$ and $y_R=g(L)$, and $y_L$ and $y_R$ are identical due to the PBC. In contrast, the after-quench Euclidean time evolution happens in a cylinder $\mathcal{M}_{\text{phy}}'$ with constant circumference $y_R-y_L$ in the $w$ coordinate. Therefore, if $\mathcal{M}_{\text{gs}}'$ and $\mathcal{M}_{\text{phy}}'$ match within a finite Euclidean time interval $(-\tau_0,\tau_0)$, the time-evolution in $\mathcal{M}_{\text{phy}}'$ can be canceled by part of the path integral in $\mathcal{M}_{\text{gs}}'$ (similar to 
the case
with simple boundary effect shown in main text Fig. 1(c)-(d)). This would allow one to calculate the entanglement entropy with path integral in spacetime manifold $\mathcal{M}_{\text{gs}}'$ in $w$ coordinate, which maps back to the simple cylinder $\mathcal{M}_{\text{gs}}$ in the $z$ coordinate.

\begin{figure}[tbp]
\begin{center}
\includegraphics[width=4in]{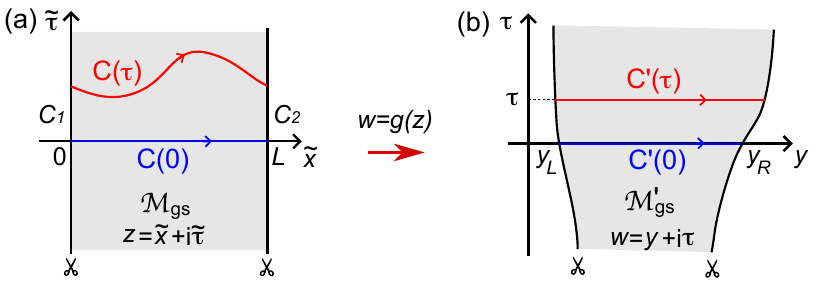}
\end{center}
\caption{
Illustration of the PBC Euclidean manifold $\mathcal{M}_{\text{gs}}'$ circumference calculation in the $z$ and $w$ coordinates, respectively.
}
\label{figSPBC}
\end{figure}

We now show that for PBC quenches with a smooth $f(x)$, the two manifolds $\mathcal{M}_{\text{gs}}'$ and $\mathcal{M}_{\text{phy}}'$ always match within a finite Euclidean time interval $(-\tau_0,\tau_0)$. This will be true if and only if $\mathcal{M}_{\text{gs}}'$ to be a cylinder with constant circumference $y_R-y_L$ within a Euclidean time interval $(-\tau_0,\tau_0)$. We denote the circumference of manifold $\mathcal{M}_{\text{gs}}'$ at a given time $\tau$ as $y_C(\tau)$. It can be rewritten as

\begin{equation}\label{contour w}
y_C(\tau)=\int dy\Big|_\tau = \oint_{C^{\prime}(\tau)} dw   
\end{equation}
where $C^{\prime}(\tau)$ is the closed constant $\tau$ contour shown in \cref{figSPBC}(b). Note that in \cref{figSPBC}(b), the two boundaries of the manifold $\mathcal{M}_{\text{gs}}'$ (grey region) are images of the vertical straight lines $\tilde{x}=0$ and $\tilde{x}=L$ in $z$ coordinate (\cref{figSPBC}(a)), and the two boundaries should be understood as identified due to PBC. Assume the constant $\tau$ contour $C^{\prime}(\tau)$ in \cref{figSPBC}(b) maps back to closed contour $C(\tau)$ in \cref{figSPBC}(a) under the inverse mapping $z=g^{-1}(w)$. With  $dw=\frac{dz}{f(z)}$, \cref{contour w} is transformed into a contour integral in the $z$ coordinates
\begin{equation}\label{contour z}
y_C(\tau)= \oint_{C(\tau)} \frac{dz}{f(z)}\ .  
\end{equation}
Because $f(x)$ is analytical and non-negative on the real axis $x\in[0,L]$, $1/f(z)$ at least has no poles and thus zero residue within a finite range of imaginary time interval $(-\tau_0,\tau_0)$. Therefore, $\tau\in(-\tau_0,\tau_0)$, the contour integral of $1/f(z)$ along closed loop $C(0)+C_{2}-C(\tau)-C_{1}$ in \cref{figSPBC}(a) is zero, where $C(0)$ is the contour $C(\tau)$ at $\tau=0$. Note that $C_{2}$ and $C_{1}$ are identical due to PBC and cancel, and $z=x$ at time $\tau=0$,  
we find
\begin{equation}
y_C(\tau)= \oint_{C(\tau)} \frac{dz}{f(z)}=\oint_{C(0)} \frac{dz}{f(z)}=\int_0^L \frac{dx}{f(x)}=y_R-y_L\ .
\end{equation}
We have thus proved that PBC guarantees that $\mathcal{M}_{\text{gs}}^{\prime}$ and $\mathcal{M}_{\text{phy}}^{\prime}$ have the same constant circumference $y_R-y_L$ for all the $\tau$ within a finite Euclidean time interval $[-\tau_0,\tau_0]$. In this case, the path integral reduces to a Euclidean path integral in the simple cylinder geometry $\mathcal{M}_{\text{gs}}$ in $z$ coordinate. 
This is similar to 
the case of simple boundary effect,
which reduces to a path integral in a simple strip geometry in $z$ coordinate. 
We have thus shown that the previous solution for CFT quenches with PBC \cite{Moosavi_2021} always corresponds to simple Euclidean path-integral spacetime geometries, in a sense similar to the simple boundary effect for CFT quenches with 
boundaries
we identified in this paper. 

\subsection{VI. 
Examples of 
inhomogeneous quenches with boundaries
and a special discussion of the M\"obius quench}

\begin{figure}[tbp]
\begin{center}
\includegraphics[width=6.8in]{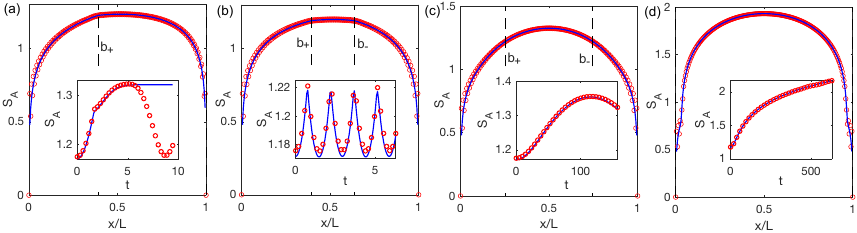}
\end{center}
\caption{
More numerical comparison examples between the CFT formula in main text Eq. (10) (blue lines) and the free-fermion tight-binding calculation (red circles), for system size $L=200$. Each panel shows $S_A(x,t_0)$ at a fixed time $t=t_0$, and the inset shows $S_A(x_0,t)$ at a fixed position $x=x_0$. The quenches and parameters are: (a) tEH with $L_1=0.3L$, $L_2=0$, $t_0=\frac{9\pi}{20}$, $x_0=0.5L$. (b) tSRD with $L_1=0.3L$, $L_2=0.2L$, $t_0=\frac{9\pi}{20}$, $x_0=0.5L$. (c) M\"obius with $\lambda=0.5$, $t_0=25\pi$, $x_0=0.5L$. (d) SSD (M\"obius with $\lambda=1$), $t_0=100\pi$, $x_0=0.5L$.
}
\label{figSI}
\end{figure}

In this section, we give calculation details and additional discussions on the quench examples we considered in main text Tab. I, and show more plots of comparison between the CFT formula and the tight-binding numerical calculations in \cref{figSI}.

In the main text, we have defined the light cone coordinates $x_\pm\in[0,L]$ as initial positions of a quasiparticle reaching point $(x,t)$ from the left/right, which we have rewritten in \cref{reflected-w*}.

Once we have $y=g(x)$ for $x\in[0,L]$, for a given time $t$, we can calculate $y_\pm$ and $x_\pm$ from \cref{reflected-w*}, and calculate
\begin{equation}
    \frac{\partial x_\pm}{\partial x}=\frac{\partial x_\pm}{\partial y_\pm}\frac{\partial y_\pm}{\partial x}=\frac{f(x_\pm)}{f(x)}\ .
\end{equation}
Then we have all the ingredients to calculate the entanglement entropy in the main text Eq. (10). So we will only specify $y=g(x)$ for $x\in[0,L]$ for our quench examples shown in the main text Tab. I. The expression for $S_{A}$ comes from substituting $\frac{\partial x_\pm}{\partial x}$ and $x_\pm=g^{-1}(y_\pm)$ in the main text Eq. (10).

1. \emph{Truncated entanglement Hamiltonian (tEH) quench}. The entanglement quench is the quench associated with the deformation:
\begin{equation}
\begin{split}
f(x)=\frac{\left(x+L_1\right)\left(L+L_2-x\right)}{L+L_1+L_2}\ ,
\end{split}
\end{equation}
which yields
\begin{equation}
\begin{split}
g(x) =\int \frac{d x}{f(x)}=\ln \left(\frac{x+L_1}{L+L_2-x}\right) 
\end{split}
\end{equation}
Unless $L_1=L_2=0$, the tEH quench has generic boundary effect.

Additional to the main text, \cref{figSI}(a) shows the free fermion numerics of tEH quench with $L_1=0.3L$ and $L_2=0$: Compared with the main text Fig. 1(a), there is only one hot spot $x_{h}=L$ in this case, and $S_A(x,t)$ from the CFT formula saturates at large $t$. $S_A(x,t)$ from tight-binding calculations deviates from the CFT formula at large $t$, as the spatial UV cutoff (lattice constant) around the hot spot prevents the saturation.

2. \emph{Truncated SRD (tSRD) quench}. The tSRD quench is the quench associated with the truncated square root deformation with 
\begin{equation}f(x)=\sqrt{\left(x+L_1\right)\left(L+L_2-x\right)}\ .
\end{equation}
This gives
\begin{equation}
g(x) =\int \frac{d x}{f(x)}
\quad \rightarrow \quad \frac{x+L_1}{L_0}=\frac{1-\cos g(x)}{2}\quad \rightarrow \quad g(x)=\cos^{-1}{\left(\frac{L+L_2-L_1-2x}{L+L_1+L_2} \right)}\ .
\end{equation}
The tSRD quench always has generic boundary effect.

3. \emph{The rainbow quench}. The rainbow quench we consider here is the quench associated with the deformation 
\begin{equation}
f(x)=e^{-k x}\ ,\qquad x\in[0,L]\ . 
\end{equation}
Here we take $L$ finite for the purpose of comparison with the tight-binding numerical results (which can only be done for finite $L$). Taking the $L\rightarrow\infty$ limit yields the rainbow quench with $x \in[0,+\infty)$. This corresponds to
\begin{equation}
g(x) =\int \frac{d x}{f(x)}=\frac{e^{k x}}{k}\ .
\end{equation}
It has generic boundary effect. In thel limit $L\rightarrow \infty$, the rainbow quench always has $n_-=0$, and $n_+\le 1$.

4. \emph{The M\"obius quench}. The M\"obius quench is the quench associated with the deformation 
\begin{equation}
f(x)=1-\lambda \cos\left(\frac{2\pi x}{L}\right)\ . 
\end{equation}
By defining $L_\text{eff}=\frac{L}{\sqrt{1-\lambda^2}}$ and $a=\frac{\sqrt{1-\lambda^2}}{1+\lambda}$, one derives the relation between $x$ and $y=g(x)$:
\begin{equation}\label{seq-mobius}
y=g(x)=\int \frac{dx}{f(x)}\ ,\quad \rightarrow \quad e^{i \frac{2 \pi x}{L}}
=\frac{(1+a) e^{i \frac{2\pi y}{L_\text{eff}}}-(1-a)}{(1-a) e^{i \frac{2\pi y}{L_\text{eff}}}-(1+a)}\ ,
\end{equation}
From which we have 
\begin{equation}
g(x)=\frac{L_{\text {eff }}}{\pi} \tan ^{-1}\left(-\frac{a}{\tan \frac{\pi x}{L}}\right)\ .
\end{equation}
The range of $y=g(x)$ is $y\in \left[-\frac{L_\text{eff}}{2},\frac{L_\text{eff}}{2}\right]$.

The M\"obius quench has simple boundary effect, which can be seen from the matching Euclidean spacetime boundaries $\partial \mathcal{M}_{\text{phy}}^{\prime}$ and $\partial \mathcal{M}_{\text{gs}}^{\prime}$ at small Euclidean time $\tau$. The boundary $\partial \mathcal{M}_{\text{phy}}^{\prime}$ is always $y=\pm L_\text{eff}/2$ and $\tau\in\mathbb{R}$. The boundary $\partial \mathcal{M}_{\text{gs}}^{\prime}$ is mapped via $w=g(z)$ from the straight boundaries $z=i\tilde{\tau}$ and $z=L+i\tilde{\tau}$.  
From \cref{seq-mobius}, and using the fact that $e^{i \frac{2\pi z}{L}}=e^{-\frac{2\pi \tilde{\tau}}{L}} >0$ with $\tilde{\tau} \in \mathbb{R}$, one finds that the boundary $\partial \mathcal{M}_{\text{gs}}^{\prime}$ in coordinate $w=y+i\tau$ is located at three $y$ positions:

(1) $y=-L_\text{eff}/2$, $\tau\in\mathbb{R}$\ .

(2) $y=L_\text{eff}/2$, $\tau\in\mathbb{R}$\ .

(3) $y=0$, $|\tau|\ge \tau_0$ with $\tau_0=\frac{L_\text{eff}}{2\pi} \ln{(\frac{1+a}{1-a})}$\ .

The SSD quench case is given by taking the limit $\lambda\rightarrow 1$, in which case $\tau_{0} \rightarrow \frac{L}{2\pi}$.

Therefore, the boundaries in $w$ for the M\"obius quench is as shown the main text Fig. 1(c), and the two boundaries $\partial \mathcal{M}_{\text{phy}}^{\prime}$ and $\partial \mathcal{M}_{\text{gs}}^{\prime}$ match within the Euclidean time interval $\tau\in(-\tau_0,\tau_0)$, and simple boundary effect applies. Thus, the previous results \cite{goto2023scrambling,Lapierre_2021,Fan_2020} relying on analytical continuations agree with ours. Comparison of the entanglement entropy $S_A$ with the numerical tight-binding calculation for M\"obius and SSD quenches are shown in \cref{figSI}(c)-(d), which shows analyticity at all $x\in(0,L)$. In the SSD case, $S_A(x,t)$ from the CFT formula is in a heating phase with perpetual linear growth in $t$, because of the two hot spots $x=0$ and $x=L$, as explained in the main text.

5. \emph{The half M\"obius quench}. Only in the case of half M\'obius quench, which has
\begin{equation}
f(x)=1-\lambda \cos\left(\frac{\pi x}{L}\right), 
\end{equation}
the boundaries $\partial \mathcal{M}_\text{gs}^{\prime}$ and $\partial \mathcal{M}_\text{phy}^{\prime}$ match entirely in the whole spacetime region. In this case,
\begin{equation}\label{seq-mobius}
y=g(x)=\int \frac{dx}{f(x)}\ ,\quad \rightarrow \quad e^{i \frac{\pi x}{L}}=\frac{(1+a) e^{i \frac{\pi y}{L_\text{eff}}}-(1-a)}{(1-a) e^{i \frac{\pi y}{L_\text{eff}}}-(1+a)}\ 
\end{equation}
where $L_\text{eff}=\frac{L}{\sqrt{1-\lambda^2}}$ and $a=\frac{\sqrt{1-\lambda^2}}{1+\lambda}$. The boundary $\partial \mathcal{M}_{\text{phy}}^{\prime}$ is always $y=0$ and $y=L_\text{eff}$ with $\tau\in\mathbb{R}$. The boundary $\partial \mathcal{M}_\text{gs}^{\prime}$ is given by $w=g(z)$ from $z=i\tilde{\tau}$ or $z=L+i\tilde{\tau}$ with $\tilde{\tau}\in\mathbb{R}$, which correspond to $e^{i \frac{\pi z}{L}}=\pm e^{-\frac{\pi \tilde{\tau}}{L}} \in \mathbb{R}$. Therefore, $e^{i \frac{\pi w}{L_\text{eff}}}$ can be any real number, corresponding to the boundaries $w=i \tau$ or $L_\text{eff}+i \tau$ for any $\tau \in \mathbb{R}$. In this case, the boundaries match entirely, and there is no quench dynamics.

The simple boundary effect for the 
M\"obius quench and half-M\"obius quench can also be understood in our ancillary mirror PBC quench problem picture as follows. 
The M\"obius quench corresponds to an ancillary mirror PBC problem with double spatial frequency M\"obius quench, which can be viewed as generated by the Virasoro generators $L_0$ and $L_{\pm2}$, which has nontrivial quantum dynamics. In contrast, the half-M\"obius quench corresponds to ancillary PBC problem of the usual M\"obius quench, which can be generated by the Virasoso generators $L_0$ and $L_{\pm1}$. Since $L_0$ and $L_{\pm1}$ are global conformal generators and do not change the PBC ground state, quantum dynamics in this case is absent.

\subsection{VII. System size effect and filling effect}

\begin{figure}[tbp]
\begin{center}
\includegraphics[width=6.0
in]{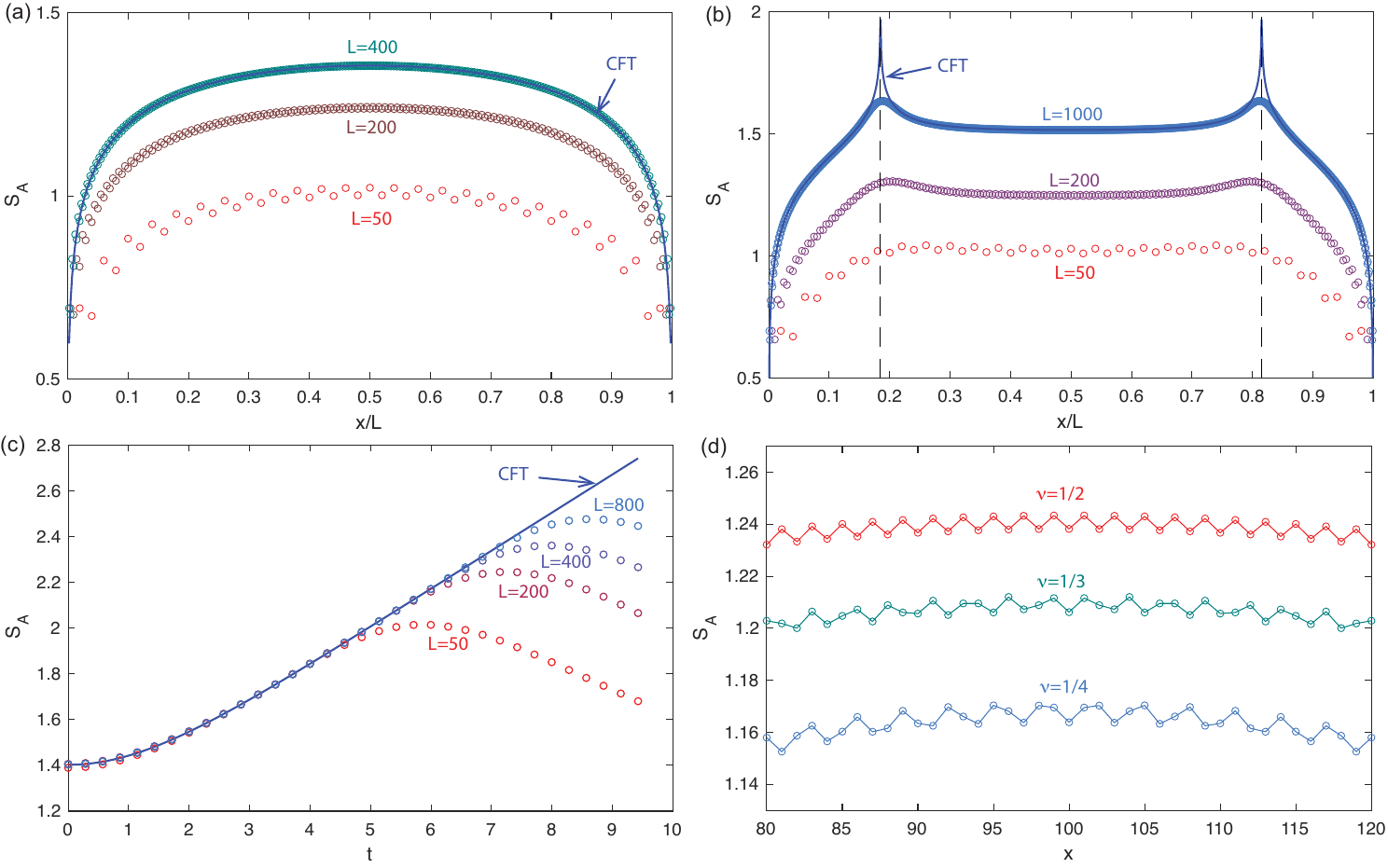}
\end{center}
\caption{More numerical comparison examples between the CFT formula in main text Eq. (10) (lines) and the free-fermion tight-binding calculation (circles), for different system sizes $L$ (labeled in the figure) in (a)-(c) and for different fillings in (d). (a) and (b) show $S_A(x,t_0)$ as a function of $x$ at a fixed time $t=t_0$; the parameters are given by (a) EH (tEH with $L_1=L_2=0$) quench, $t_0=\frac{9\pi}{20}$. (b) SRD (tSRD with $L_1=L_2=0$) quench, $t_0=\frac{3\pi}{10}$. The quenches are defined in the main text Tab. I. (c) shows $S_A(x_0,t)$ (with an offset added to collapse all the curves to the same point at $t=0$) as a function of time $t$ at a fixed position $x=x_0$, for EH quench at a fixed position $x=x_0=0.5L$. (d) shows the zoom-in numerical tight-binding lattice model $S_A(x,t_0)$ at different filling factors $\nu$ for EH quench with $L=200$, $t_0=\frac{9\pi}{20}$. Compared to the CFT formula, the numerical $S_A$ exhibits a small oscillation with respect to $x$, where the oscillation period depends on filling $\nu$ (which is UV physics not captured by the CFT result).
}
\label{fignew}
\end{figure}

 \cref{fignew} shows the effect of system size $L$ and filling factor 
(number of fermions per site) $\nu$ on entanglement entropy of the tight-binding model quench numerical calculations. In the main text, we have chosen $L=200$. By increasing the number of lattice sites $L$, our numerical result approaches the CFT prediction in main text Eq. (10). In \cref{fignew}(a)-(b) which shows $S_A$ at a constant time $t=t_0$ with respect to $x$, the relative shift between different $L$ can be understood by the constant term $\frac{c}{6}\ln (L/\epsilon)$ in the CFT prediction in main text Eq. (10) (only the CFT predicted curve simulating the largest $L$ numerical result is plotted as solid lines). In \cref{fignew}(c) which shows $S_A$ with respect to time $t$, the numerical results for different system sizes $L$ are added with an offset to collapse them to the same value at time $t=0$, for convenience of comparison between them.

To summarize, the UV cutoff induced deviation from the CFT prediction is a small oscillation as a function of site $x$: the oscillation amplitude decreases as $L$ increases as in \cref{fignew} (a), (b) and (c), and the oscillation period depends on the fermion filling $\nu$ per site (the period is 2 when the filling factor is $\nu=1/2$) as in \cref{fignew} (d). 

In the case where the system has a hot spot, the CFT predicted entanglement entropy diverges linearly as a function of time $t$; as \cref{fignew} (c) shows, the tight-binding model result follows this behavior at early time $t$ up to certain time bounded by the log of the Hilbert space dimension, which increases as $L$ increases. In most cases, the inifinite entanglement entropy in time at most grows linearly in $t$ towards infinity. The divergence corresponds to a heating phase with perpetual linear growth in $t$, because there exist hot spots. As we already explained in the main text, a hot spot $x_h\in[0,L]$ is where $f(x)\propto|x-x_h|^{\eta}$ with $\eta\ge1$, thus the time $\int^{x_h}\frac{dx}{f(x)}$ for particles to reach $x_h$ diverges, and heat (entropy) is trapped at $x_h$. $S_A(x,t)$ from tight-binding calculations is eventually upper bounded at large $t$, as the spatial UV cutoff (lattice constant) around the hot spots prevents the time divergence. In the examples we considered in Table I in the main text, the hotspots always have $\eta=1$. In \cref{fignew} (c), we have hotspots for EH at $x=0$ and $L$, where $f(x) \simeq x$ and $L-x$, respectively. In the lattice model with the UV cutoff being the lattice constant 1, the time for the particles to reach within the UV cutoff of the hotspot is $\int_1^{L}\frac{dx}{f(x)} \simeq \ln L$. This is the time after which the divergence of entanglement entropy stops. For $L=200$, this time is about 5.3, which agrees well with \cref{fignew} (c).

A special case is the square root deformation (without truncation) in \cref{fignew} (b), in which the entanglement entropy from our CFT formula diverges at $b_\pm$ at finite time $t$. The boundary points $x_0=0$ or $L$ has $f(x)\propto|x-x_0|^{1/2}$, so these points are not hot spots, and the modes from the two boundary points $x_0$ can propagate to $b_\pm$ after a finite time $t$. In this case, the understanding of the divergence of entanglement entropy is as follows. First, it is known that in CFT, the entanglement entropy $S_A$ is divergent unless a UV cutoff $\epsilon$ is introduced, yielding for instance $S_A\sim \ln(x/\epsilon)$ for an interval of length $x$. Heuristically, this is ignoring all the entangled pairs of particles within size (distance between the particles) $\epsilon$. In our case, $f(x)$ plays the role of local speed of light, which reaches zero at the boundary points $x_0=0$ or $L$. When an entangled pair of size smaller than $\epsilon$ at the boundary points $x_0$ propagates to $x=b_\pm$, its size will be amplified by the ratio of the speed of light at $b_\pm$ to that at $x_0$, and this ratio is $\frac{f(b_\pm)}{f(x_0)}=\infty$. Thus, there are infinitely many entangled pairs, whose size were smaller than $\epsilon$, now become large size entangled pairs across $b_\pm$, which yields a diverging entanglement entropy at $b_\pm$. In simpler words, this divergence results from the infinite contraction ratio of the UV cutoff scale $\epsilon$ (equivalent to the ratio of speed of light), which corresponds to the factor $|\partial x_\pm/\partial x|^{-1/2}$ in our formula (Eq. (10) of the main text). In lattice model calculations, the numerical entanglement entropy is never divergent, but only aproaches this diverging behavior in the thermodynamic limit ($L\rightarrow \infty$), as shown in \cref{fignew} (b). 

\end{document}